\documentclass[11pt]{article}
\usepackage{amssymb,amsmath,amsfonts}
\usepackage{graphicx}
\usepackage{graphics}
\usepackage{eepic,epsfig}

\@addtoreset{equation}{section}

\makeatother

\begin{document}

\title{Vacuum effects induced by a plate in de Sitter spacetime in the presence of a cosmic string}
\author{W. Oliveira dos Santos$^{1}$\thanks{%
E-mail: wagner.physics@gmail.com}  \ and E. R. Bezerra de Mello$^{1}$\thanks{%
E-mail: emello@fisica.ufpb.br} \\
\textit{$^{2}$Departamento de F\'{\i}sica, Universidade Federal da Para\'{\i}%
ba}\\
\textit{58.059-970, Caixa Postal 5.008, Jo\~{a}o Pessoa, PB, Brazil}\vspace{%
0.3cm}}
\maketitle

\begin{abstract}
In this paper we investigate the vacuum expectation values of the field squared and the energy-momentum tensor associated to a charged massive scalar quantum field in a  $(1+D)$-dimensional de Sitter spacetime induced by a plate (flat boundary) and a carrying-magnetic-flux cosmic string.  In our analysis we admit that the flat boundary is perpendicular to the string, and  the scalar field obeys the Robin boundary condition on the plate. In  order to do develop this analysis, we obtain the complete set of normalized positive-energy solution of the Klein-Gordon equation compatible with the model setup. Having obtained these bosonic modes, we construct the corresponding Wightman function. The latter is given by the sum of two terms: one associated with the boundary-free spacetime, and the other induced by the flat boundary. Although we have imposed the Robin boundary condition on the field, we apply our formalism considering specifically the Dirichlet and Neumann boundary conditions. The corresponding parts have opposite signs. Because the analysis of bosonic vacuum polarization in boundary-free de Sitter space and in presence of a cosmic string, in some sense, has been developed in the literature, here we are mainly interested in the calculations of the effects induced by the boundary. In this way, closed expressions for the corresponding expectation values are provided, as well as their asymptotic behavior in different limiting regions. We show that the conical topology due to the cosmic string enhances the boundary induced vacuum polarization effects for both field squared and the energy-momentum tensor, compared to the case of a boundary in pure de Sitter spacetime.  Moreover, the presence of cosmic string and boundary induce non-zero stress along the direction normal to the boundary. The corresponding vacuum force acting on the boundary is also investigated. 
\end{abstract}


\bigskip

\section{Introduction}
\label{Int}
De Sitter (dS) spacetime is one of the most interesting solutions allowed by the Theory of General Relativity. This curved maximally symmetric spacetime allows that several physical problems in Quantum Field Theory can be exactly solvable; moreover,  the importance of these theoretical analysis increased by the appearance of the inflationary cosmology scenario \cite{Linde}. In various inflation models, the expansion is taken by an approximately dS spacetime sourced by the potential energy of a scalar field, called inflaton. This short period of expansion naturally provides solution to several challenging issues within the standard Cosmology, such as the horizon and flatness problems for instance. Furthermore, quantum fluctuations in the inflaton field introduces small inhomogeneities which may have important consequences on the formation of large scale structures of the Universe at late stages. On the experimental grounds, recent astronomical observations of high redshift supernovae, galaxy clusters and cosmic microwave background \cite{Ries2007, Spergel2007, Seljak2006} suggest that the Universe is currently under accelerating expansion and could be approximately described by a world with a positive cosmological constant.

Cosmic strings are linear topologically stable gravitational defects which may have been created in the early Universe after Planck time by a vacuum phase transition \cite{Kibble,V-S}. The gravitational field produced by a cosmic string may be approximated by a planar angle deficit in the surface orthogonal to the string. The simplest theoretical model which describes a straight and infinitely long cosmic string is given by a Dirac-delta type distribution for the energy-momentum tensor along the linear defect. This object can also be described by coupling the energy-momentum tensor associated with the Maxwell-Higgs system \cite{N-O} with the Einstein equations \cite{Garfinkle,Linet}.

Most of the current candidates for the final fundamental theory of nature have extra dimensions as a common feature in their formulation. Topological defects were originally analyzed in 4-dimensional spacetimes \cite{Kibble,V-S} and later considered in higher dimensional schemes, such as braneworld models. In this context, the defects are supposed to live in a higher $D$-dimensional submanifold embedded in a $(4+D)$-dimensional bulk. For the particular case of a cosmic string, two additional extra dimensions are required. In this scenario, the gravitational effects due to global cosmic strings were analyzed in \cite{Cohen,Ruth} and considered as responsible for the compactification from six to four dimensions, naturally leading to the observed hierarchy between electroweak and gravitational forces. The investigations of quantum effects for different spin fields on the dS geometry were conducted by various authors (for instance, see \cite{Cher68}-\cite{Mello} and references therein). Additional effects due to cosmic strings were studied for scalar \cite{scalar}-\cite{scalar4} and fermionic fields \cite{ferm}-\cite{Spinelly}.

 Another type of vacuum polarization is due to the presence of boundaries. This is the Casimir effect. In this sense, the analysis of the Casimir effect of the scalar, vector and fermionic fields in the idealized cosmic string spacetime, obeying boundary conditions have been analyzed in \cite{BezerradeMello:2006df}, \cite{Brevik:1994ub} and \cite{BezerradeMello:2008zd}, respectively. Moreover, the Casimir effect induced by two parallel cosmic strings was investigated in \cite{Bordag1990}-\cite{Grats1}. Considering a flat boundary orthogonal to the string, the vacuum effects associated with scalar and fermionic fields have been investigated in \cite{BezerradeMello:2011sm,BezerradeMello:2012ajq}.

The analysis of the vacuum polarization associated with a real bosonic field in dS spacetime in the presence of a cosmic string has been developed in \cite{Mello_09}. The fermionic vacuum polarization in the same setup model has been considered in \cite{Mello_10}. Our objective in this paper is to extend the analysis of bosonic quantum vacuum considering charged field in this background and the presence of a flat boundary, throughout the paper also called {\it plate}. 

The plan of the paper is the following. In the next section, we present the setup model under consideration, and the complete set of normalized positive energy solutions of the Klein-Gordon equation that obeys Robin boundary conditions on a plate orthogonal to the string. Having obtained the complete set of bosonic modes, in section \ref{Wigh_func} we construct the positive frequency Wightman function. We show that this function is given by the sum of two distinct contributions: the first one associated with the $(1+D)-$dimensional dS spacetime in the presence of a cosmic string without boundary, and the other induced by the boundary condition satisfied by the scalar field. Both parts are presented in a closed form. Although we have obtained the complete set of bosonic modes satisfying Robin boundary condition on the plate, in section \ref{phi_2} we calculate the vacuum expectation value (VEV) of the field squared, considering as direct applications of the formalism, the Dirichlet and Neumann boundary conditions, separately. As we will see this VEV is given in terms of a pure dS spacetime contribution, plus a term induced by the magnetic-flux-carrying cosmic string, and finally a contribution induced by the flat boundary orthogonal to the string. Moreover, in this section we explicitly detail the asymptotic behavior of the boundary induced part in different regions. To complete the analysis we present plots of the VEV of the field squared as function of  relevant physical variables considering different values of the parameter associated with the conical structure generated by the cosmic string. In section \ref{EMT} we calculate the VEV of the energy-momentum tensor induced by the boundary considering directly the Dirichlet and Neumann conditions. We analyze specifically the energy-density and its asymptotic expressions for different regions. Also in this section we present some plots for the energy-density as function of some physical relevant variables. We finalize section \ref{EMT} by analyzing the vacuum force acting on the boundary. Finally the main results of the paper are summarized in section \ref{conc}. 

\section{Background geometry and matter field content}
\label{geom}
The main objective of this section is to obtain the complete set of positive energy solution of the Klein-Gordon equation, in a $(1+D)$-dimensional dS spacetime in the presence of an infinitely long straight cosmic string, obeying the Robin boundary condition on a flat boundary orthogonal to the string. In order to do that, we write the corresponding line element in cylindrical coordinate \cite{Mello_09,Mello_10}:
\begin{equation}
	\label{ds1}
	ds^2=dt^2-e^{2t/a}\left(dr^2+r^{2}d\phi^2+dz^{2}+\sum_{i=1}^{D-3}dx_{i}^{2}\right) \ ,
\end{equation}
where $r\ge0$ and $\phi\in[0,2\pi/q]$. The parameter $q\geq1$ encodes the conical geometry. The coordinates  $(t,z,x_{i})\in(-\infty,\infty)$, and $a$ stands for the length scale of dS spacetime and it is related with the cosmological constant, $\Lambda$, and the curvature scalar, $R$, by the following relations:
\begin{equation}
	\Lambda=\frac{D(D-1)}{2a^2}, \quad R=\frac{D(D+1)}{a^2} \ .
\end{equation}
Besides the synchronous time coordinate above, for convenience of the discussion we adopt the conformal time, $\tau$, defined as $\tau=-a e^{-t/a}$ with $\tau\in(-\infty,0]$. By doing so, we get
\begin{equation}\label{ds2}
	ds^2=\left(\frac{a}{\tau}\right)^2%
	\left(d\tau^2-dr^2-r^{2}d\phi^2-dz^{2}-\sum_{i=1}^{D-3}dx_{i}^{2}\right) \ .
\end{equation}
Note that the line element inside brackets corresponds to the metric tensor associated with  an idealized cosmic string in Minkowski spacetime.

In this paper, we want to analyze the vacuum effects associated to a propagating charged scalar field in the dS spacetime in the presence of a magnetic-carrying-flux cosmic string and a flat boundary perpendicular to the string. For this purpose we consider the following Klein-Gordon field equation:
\begin{equation}\label{KG}
	(g^{\mu\nu}\mathcal{D}_{\mu}\mathcal{D}_{\nu}+m^2+\xi R)\varphi(x)=0 \ ,
\end{equation}
where $\mathcal{D}_{\mu}=\nabla_{\mu}+ieA_{\mu}$ with $m$ being the mass of the field. Moreover, in the equation above we have introduced a non minimal coupling between the background curvature and the scalar field through the term $\xi R$, with $\xi$ being the curvature coupling constant and $R$ the Ricci scalar. The magnetic flux along the string axis is implemented through the gauge field $A_{\mu}=A_{\phi}\delta_{\mu}^{\phi}$, with a constant $A_{\phi}=-q\Phi/2\pi$, being $\Phi$ the magnetic flux along the string.

For consideration of the single flat boundary we impose that solutions of the Klein-Gordon equation \eqref{KG} obey the Robin boundary conditions given by
\begin{equation}
	(1+\beta n^{\mu}\nabla_{\mu})\varphi(x)\Big|_{z=0}=0 \ ,
	\label{RBC}
\end{equation}
where $\beta$ is a constant coefficient. In particular, for $\beta=0$ and $\beta=\infty$, the Robin boundary conditions are reduced to the Dirichlet and Neumann boundary conditions, respectively. In both sides of the plate one has $n^{\mu}=\delta_{z}^{\mu}$. According to the notation above, the flat boundary is located at $z=0$. Moreover, note that in our setup problem the cosmic string is perpendicular to the boundary.

\subsection{Bosonic modes}
\label{Sec2}
In this section our goal is to find the complete set of normalized positive energy solutions of the Klein-Gordon equation \eqref{KG} that satisfies \eqref{RBC}. 

In the geometry of the spacetime under consideration \eqref{ds2} and with the gauge field $A_{\mu}=A_{\phi}\delta_{\mu}^{\phi}$, the Klein-Gordon equation reduces to
\begin{eqnarray}\label{EQM}
	&&\Bigg[\frac{\partial^2}{\partial\tau^2}%
	+ \frac{(1-D)}{\tau}\frac{\partial}{\partial\tau}%
	+\frac{D(D+1)\xi+(ma)^2}{\tau^2}%
	-\frac{\partial^2}{\partial r^2}%
	-\frac{1}{r}\frac{\partial}{\partial r}%
	-\frac{1}{r^2}\left(\frac{\partial}{\partial\phi}+ieA_{\phi}\right)^{2}%
	\nonumber\\
	&-&\frac{\partial^2}{\partial z^2}%
	-\sum_{i=1}^{D-3}\frac{\partial^2}{\partial(x^{i})^2}\Bigg]\varphi(x)=0 \ .
\end{eqnarray}

Considering the cylindrical symmetry of the problem, and the structure of the differential equation \eqref{EQM}, we adopt that the corresponding solutions can be presented as the variable separation as shown bellow:
\begin{equation}\label{Ansatz}
	\varphi(x)=f(\tau)R(r)h(z)e^{iqn\phi+i\vec{k}\cdot \vec{x}_{||}} \ ,
\end{equation}
where $n$ is the integer quantum number associated to the angular momentum in the azimuthal direction, $\vec{x}_{||}$ represents the set of coordinates along the $(D-3)$ extra dimensions and $\vec{k}$ the corresponding momenta. The unknown function $h(z)$ will be specified by the Robin boundary conditions obeyed by the scalar field on the plate placed at $z=0$.

Taking \eqref{Ansatz} into \eqref{EQM}, we get
\begin{equation}\label{DE-z}
	\frac{\partial^2 h(z)}{\partial z^2}=-k_{z}^{2}h(z) \ ,
\end{equation}
and the following differential equations for the functions $f(\tau)$ and $R(r)$:
\begin{equation}\label{DE-tau}
	\Bigg[	\frac{\partial^2}{\partial\tau^2}+\frac{(1-D)}{\tau}\frac{\partial}{\partial\tau}+\frac{D(D+1)\xi+(ma)^2}{\tau^2}+\lambda^{2}\Bigg]f(\tau)=0 \ ,
\end{equation}
and
\begin{equation}\label{DE-r}
	\Bigg[\frac{\partial^2}{\partial r^2}+\frac{1}{r}\frac{\partial}{\partial r}-\frac{q^{2}(n+\alpha)^2}{r^2}+p^{2}\Bigg]R(r)=0 \ ,
\end{equation}
with
\begin{equation}\label{DR}
	\lambda=\sqrt{p^2 + k_{z}^{2} + \vec{k}^2 }
\end{equation}
and the notation
\begin{equation}
	\alpha=\frac{eA_{\phi}}{q}=-\frac{\Phi}{\Phi_{0}} \ ,
	\label{alpha}
\end{equation}
being $\Phi_{0}=2\pi/e$ the quantum flux.

The solution for \eqref{DE-r}  regular at $r=0$ is given by
\begin{equation}\label{sol-r}
	R(r)=J_{q|n + \alpha|}(pr)  \  ,
\end{equation}
being $J_{\mu}(x)$ the Bessel function of first kind \cite{Grad}. The solution for the time-dependent equation is given by the linear combination of Hankel functions:
\begin{equation}\label{sol-tau}
	f(\tau)=\eta^{D/2}(c_1 H_{\nu}^{(1)}(\lambda \eta) + c_2 H_{\nu}^{(2)}(\lambda \eta)) \ ,
\end{equation}
with the order given by
\begin{equation}
	\nu=\sqrt{D^2/4 - m^2a^2 - \xi D(D+1)} \ .
	\label{nu}
\end{equation}
Moreover, in \eqref{sol-tau} we have introduced the variable $\eta=|\tau|$ and $H_{\nu}^{(l)}(x)$, $l=1, \ 2$, represents the Hankel function \cite{Grad}. Different choices of the coefficients $c_{1,2}$ in \eqref{sol-tau} lead to different choices of the vacuum state. In fact, there is a one-parameter family of de Sitter invariant vacua, named $\alpha$-vacua \cite{Mottola1984,Allen1985}. The Bunch-Davies vacuum corresponds to $c_2=0$. The short-distance behaviour of the corresponding two-point functions has the same structure as in flat spacetime. Moreover, as pointed out in Ref. \cite{Goldstein2002}, all $\alpha$-vacuum states lead to infinite energy density at the end of inflationary expansion, with the exception of the Bunch-Davies vacuum. Therefore, in this paper we adopt the Bunch-Davies vacuum.

As to the solution of the $z$-dependent equation, obeying the Robin boundary conditions \eqref{RBC} on the flat boundary at $z=0$, is given by
\begin{equation}\label{sol-z}
	h(z)=\cos[k_{z}z+\alpha_{1}(k_z)] \ ,
\end{equation}
with the notation
\begin{equation}
	e^{2i\alpha_{1}(k_z)}=\frac{i\beta k_z-1}{i\beta k_z+1} \ .
	\label{FBC}
\end{equation}

Finally, combining \eqref{sol-r}, \eqref{sol-tau} and \eqref{sol-z}, the mode functions that obey the Klein-Gordon equation \eqref{KG} and also satisfy the Robin boundary conditions \eqref{RBC} on the plate is given by
\begin{equation}\label{MF}
	\varphi_{\sigma}(x)=C_{\sigma}\eta^{D/2}H_{\nu}^{(1)}(\lambda \eta)J_{q|n + \alpha|}(pr)\cos[k_{z}z+\alpha_{1}(k_z)]e^{iqn\phi+i\vec{k}\cdot \vec{x}_{||}} \ ,
\end{equation}
where $\sigma=\{\lambda, p, n, k_z, \vec{k}\}$ is the set of quantum numbers specifying each mode of the field. The coefficient $C_{\sigma}$ is determined by the orthonormalization condition
\begin{equation}\label{NCF}
	-i\int d^{D-1}x\int_{0}^{\infty}dz\sqrt{|g|}g^{00}[\varphi_{\sigma}(x)\partial_{\tau}\varphi_{\sigma^\prime}^{\ast}(x)-\varphi_{\sigma^\prime}^{\ast}(x)\partial_{\tau}\varphi_{\sigma}(x)]=\delta_{\sigma,\sigma^{\prime}} \ ,
\end{equation}
where the integral is evaluated over the spatial hypersurface $\tau=$ const, and $\delta_{\sigma,\sigma^{\prime}}$ represents the Kronecker-delta for discrete indices and Dirac-delta function for continuous ones. For the mode functions in \eqref{MF}, the normalization condition leads to
\begin{equation}\label{NC}
	|C_{\sigma}|^{2}=\frac{qpe^{i(\nu-\nu^{\ast})\pi/2}}{2(2\pi)^{D-2}a^{D-1}} \ .
\end{equation}

\section{Wightman function}
\label{Wigh_func}
Our aim in this paper is to investigate the vacuum polarization effects emerging from the background setup described in the previous section. For this purpose we will make use of the Wightman function. This function is of special interest since its form is useful in calculations of vacuum expectation values of physical observables that depend on bilinear field operators. In particular, the properties of the vacuum can be described by the Wightman function, $W(x,x^{\prime})=\langle 0|\hat{\varphi}(x)\hat{\varphi}^{\dagger}(x^{\prime}) |0\rangle$, where $|0\rangle$ stands for the vacuum state. For evaluation of this function, we will adopt the mode sum formula technique by which the Wightman function takes the form:
\begin{equation}\label{WF}
	W(x,x^{\prime})=\sum_{\sigma}\varphi_{\sigma}(x)\varphi_{\sigma}^{\ast}(x^{\prime}) \ .
\end{equation}
where $\sum_{\sigma}$ stands for summations over the set of discrete and integrals over continuous quantum numbers, $\sigma=\{\lambda, p, n, k_z, \vec{k}\}$.

Taking \eqref{MF}, along with the coefficient \eqref{NC}, into \eqref{WF}, we obtain
\begin{eqnarray}
	W(x,x^{\prime})&=&\frac{8q(\eta \eta^{\prime})^{D/2}}{(2\pi)^{D}a^{D-1}} \sum_{n=-\infty}^{\infty}e^{inq\Delta\phi}\int_{0}^{\infty} dp p J_{q|n + \alpha|}(pr)J_{q|n + \alpha|}(pr^{\prime})\int d\vec{k}e^{i\vec{k}\cdot\Delta\vec{x}_{||}}\nonumber\\%
	&\times&\int_{0}^{\infty}dk_{z}K_{\nu}(e^{-\pi i/2}\eta \lambda)K_{\nu}(e^{\pi i/2}\eta^{\prime} \lambda)\cos[k_{z}z+\alpha_{1}(k_z)]\cos[k_{z}z^{\prime}+\alpha_{1}(k_{z})] \ ,
	\label{W2}
\end{eqnarray}
where $\Delta\phi=\phi-\phi^\prime$ and $\Delta \vec{x}_{\parallel}=\vec{x}_{\parallel}-\vec{x}_{\parallel}^\prime$ and $K_\nu(x)$, represents the Macdonald function \cite{Abra}. Moreover, to obtain the expression above we have introduced the notation $\lambda=\sqrt{p^{2} + k_{z}^{2} + \vec{k}^{2}}$ and used the relation \cite{Abra}
\begin{equation}
	e^{i(\nu-\nu^{\ast})\pi/2}H_{\nu}^{(1)}(\lambda \eta)\big[H_{\nu}^{(1)}(\lambda \eta^{\prime})\big]^{\ast}=\frac{4}{\pi^2}K_{\nu}(-i\lambda\eta)K_{\nu}(i\lambda\eta^{\prime}) \ .
\end{equation}
In order to integrate over the quantum numbers $p$ and $\vec{k}$, we can make use of the following identity \cite{Watson}:
\begin{equation}
	K_{\nu}(a)K_{\nu}(b)=\frac{1}{2}\int_{-\infty}^{\infty}dy\int_{0}^{\infty}dww^{-1}\exp[-2\nu y-abw^{-1}\cosh(2y)-w/2-(a^2+b^2)/2w] \ .
\end{equation}
Taking this expression into \eqref{W2}, we get
\begin{eqnarray}
	W(x,x^{\prime})&=&\frac{4q(\eta \eta^{\prime})^{D/2}}{(2\pi)^{D}a^{D-1}} \int_{-\infty}^{\infty}dy\int_{0}^{\infty}dww^{-1}e^{-2\nu y-w/2}\nonumber\\%
	&\times&\sum_{n=-\infty}^{\infty}e^{inq\Delta\phi}\int dp p e^{-\sigma p^2/2w} J_{q|n + \alpha|}(pr)J_{q|n + \alpha|}(pr^{\prime})\int d\vec{k}e^{i\vec{k}\cdot\Delta\vec{x}_{||}-\sigma\vec{k}^2/2w}\nonumber\\%
	&\times&\int_{0}^{\infty}dk_{z}e^{-\sigma k_{z}^2/2w}\cos[k_{z}z+\alpha_{1}(k_z)]\cos[k_{z}z^{\prime}+\alpha_{1}(k_{z})] \ ,
	\label{WF3}
\end{eqnarray}
with the notation
\begin{equation}
	\varepsilon=2\eta\eta^\prime\cosh(2y)-\eta^2-\eta^{\prime2} \ .
	\label{sigma}
\end{equation}
The integrations over $p$ and $\vec{k}$ can be performed with the help of the formulas given in \cite{Grad}. By doing this, we obtain
\begin{eqnarray}
	W(x,x^{\prime})&=&\frac{4q(\eta \eta^{\prime})^{D/2}}{(2\pi)^{(D+3)/2}a^{D-1}} \int_{-\infty}^{\infty}dy\int_{0}^{\infty}dww^{-1}\left(\frac{w}{\varepsilon}\right)^{(D-1)/2}e^{-2\nu y-w/2-(r^2+r^{\prime2}+\Delta\vec{x}_{||}^{2})w/(2\varepsilon)}\nonumber\\%
	&\times&\sum_{n=-\infty}^{\infty}e^{inq\Delta\phi}I_{q|n+\alpha|}\left(\frac{rr^{\prime}w}{\sigma}\right)\int_{0}^{\infty}dk_{z}e^{-\varepsilon k_{z}^2/2w}\cos[k_{z}z+\alpha_{1}(k_z)]\nonumber\\%
	&\times&\cos[k_{z}z^{\prime}+\alpha_{1}(k_{z})] \ .
	\label{WF4}
\end{eqnarray}
Introducing a new variable of integration, $\chi=w/2\varepsilon$, we can write
\begin{eqnarray}
	W(x,x^{\prime})&=&\frac{q(\eta \eta^{\prime})^{D/2}}{\pi^{\frac{D+3}{2}}a^{D-1}} \int_{0}^{\infty}d\chi\chi^{(D-3)/2}e^{-(r^2+r^{\prime2}+\Delta\vec{x}_{||}^{2})\chi}\int_{-\infty}^{\infty}dye^{-2\nu y-\sigma\chi}\nonumber\\%
	&\times&\sum_{n=-\infty}^{\infty}e^{inq\Delta\phi}I_{q|n+\alpha|}\left(2rr^{\prime}\chi\right)\int_{0}^{\infty}dk_{z}e^{- k_{z}^2/4\chi}\cos[k_{z}z+\alpha_{1}(k_z)]\nonumber\\%
	&\times&\cos[k_{z}z^{\prime}+\alpha_{1}(k_{z})] \ .
	\label{WF5}
\end{eqnarray}
Now we can integrate over $y$. By taking into account \eqref{sigma} and using the formula \cite{Mello_09}
\begin{equation}
	\int_{-\infty}^{\infty}dye^{-2\nu y-2x\eta\eta^{\prime}\cosh(2y)}=K_{\nu}(2\eta\eta^{\prime}x) \ ,
\end{equation}
we can write \eqref{WF5} as
\begin{eqnarray}
	W(x,x^{\prime})&=&\frac{q(\eta \eta^{\prime})^{D/2}}{\pi^{\frac{D+3}{2}}a^{D-1}} \int_{0}^{\infty}d\chi\chi^{(D-3)/2}e^{-(r^2+r^{\prime2}+\Delta\vec{x}_{||}^{2}-\eta^2-\eta^{\prime2})\chi}\nonumber\\%
	&\times&\sum_{n=-\infty}^{\infty}e^{inq\Delta\phi}I_{q|n+\alpha|}\left(2rr^{\prime}\chi\right)K_{\nu}(2\eta\eta^{\prime}\chi)\int_{0}^{\infty}dk_{z}e^{- k_{z}^2/4\chi}\cos[k_{z}z+\alpha_{1}(k_z)]\nonumber\\%
	&\times&\cos[k_{z}z^{\prime}+\alpha_{1}(k_{z})] \ .
	\label{WF6}
\end{eqnarray}
We now turn to the integration over $k_z$. In this sense, we use the identity
\begin{equation}
	2\cos[k_{z}z+\alpha_{1}(k_z)]\cos[k_{z}z^{\prime}+\alpha_{1}(k_z)]=\cos[k_z(z-z^{\prime})]+ \cos[k_z(z+z^{\prime})+2\alpha_{1}(k_z)] \ ,
\end{equation}
allowing us to write the Wightman function as
\begin{equation}
	W(x,x^{\prime})=W_{\rm{cs}}^{\rm{dS}}(x,x^{\prime})+W_{\rm{pl}}(x,x^{\prime}) \ .
	\label{WF_terms}
\end{equation}
The first term on the r.h.s. of the above expression is the Wightman function due to the cosmic string with its magnetic flux in dS spacetime without boundary. Note that in the absence of the magnetic flux, this term coincides with the one investigated in Ref. \cite{Mello_09}. Explicitly, this term reads
\begin{eqnarray}
	W_{\rm{cs}}^{\rm{dS}}(x,x^{\prime})&=&\frac{q(\eta \eta^{\prime})^{D/2}}{2\pi^{\frac{D+3}{2}}a^{D-1}} \int_{0}^{\infty}d\chi\chi^{(D-3)/2}e^{-(r^2+r^{\prime2}+\Delta\vec{x}_{||}^{2}-\eta^2-\eta^{\prime2})\chi}K_{\nu}(2\eta\eta^{\prime}\chi)\nonumber\\%
	&\times&\sum_{n=-\infty}^{\infty}e^{inq\Delta\phi}I_{q|n+\alpha|}\left(2rr^{\prime}\chi\right)\int_{0}^{\infty}dk_{z}e^{- k_{z}^2/4\chi}\cos[k_{z}(z-z^{\prime})] \ .
	\label{WFcs}
\end{eqnarray}
As to the second term, it is the Wightman function induced by the plate.\footnote{The Wightman function associated to a single  plate in dS spacetime was studied in \cite{Saharian:2009ii}.} It is given by
\begin{eqnarray}
	W_{\rm{pl}}(x,x^{\prime})&=&\frac{q(\eta \eta^{\prime})^{D/2}}{2\pi^{\frac{D+3}{2}}a^{D-1}} \int_{0}^{\infty}d\chi\chi^{(D-3)/2}e^{-(r^2+r^{\prime2}+\Delta\vec{x}_{||}^{2}-\eta^2-\eta^{\prime2})\chi}K_{\nu}(2\eta\eta^{\prime}\chi)\nonumber\\%
	&\times&\sum_{n=-\infty}^{\infty}e^{inq\Delta\phi}I_{q|n+\alpha|}\left(2rr^{\prime}\chi\right)\int_{0}^{\infty}dk_{z}e^{- k_{z}^2/4\chi}\cos[k_{z}(z+z^{\prime})+2\alpha_{1}(k_z)] \ .
	\label{WFb}
\end{eqnarray}

From this point, let us develop each term separately, starting with the string-induced term \eqref{WFcs}. The integral over $k_z$ can be easily performed. After integration, \eqref{WFcs} reads
\begin{eqnarray}
	W_{\rm{cs}}^{\rm{dS}}(x,x^{\prime})&=&\frac{q(\eta \eta^{\prime})^{D/2}}{2\pi^{\frac{D}{2}+1}a^{D-1}} \int_{0}^{\infty}d\chi\chi^{D/2-1}e^{-(r^2+r^{\prime2}+\Delta\vec{x}_{||}^{2}+\Delta z^2-\eta^2-\eta^{\prime2})\chi}K_{\nu}(2\eta\eta^{\prime}\chi)\nonumber\\%
	&\times&\sum_{n=-\infty}^{\infty}e^{inq\Delta\phi}I_{q|n+\alpha|}\left(2rr^{\prime}\chi\right) \ ,
	\label{WFcs2}
\end{eqnarray}
where $\Delta z=z-z^{\prime}$.

The parameter $\alpha$ in Eq. \eqref{alpha} can be written in the form
\begin{equation}
	\alpha=n_{0}+\alpha_0, \ \textrm{with}\ |\alpha_0|<\frac{1}{2}  \  ,
	\label{const-2}
\end{equation}
being $n_{0}$ an integer number. This allow us to sum over the quantum number $n$ in \eqref{WFcs2} by using the formula obtained in \cite{deMello:2014ksa}:
\begin{eqnarray}
	&&\sum_{n=-\infty}^{\infty}e^{iqn\Delta\phi}I_{q|n+\alpha|}(x)=\frac{1}{q}\sum_{k}e^{x\cos(2\pi k/q-\Delta\phi)}e^{i\alpha_0(2\pi k -q\Delta\phi)}\nonumber\\
	&-&\frac{e^{-iqn_{0}\Delta\phi}}{2\pi i}\sum_{j=\pm1}je^{ji\pi q|\alpha_0|}
	\int_{0}^{\infty}dy\frac{\cosh{[qy(1-|\alpha_0|)]}-\cosh{(|\alpha_0| qy)e^{-iq(\Delta\phi+j\pi)}}}{e^{x\cosh{(y)}}\big[\cosh{(qy)}-\cos{(q(\Delta\phi+j\pi))}\big]} \  ,
	\label{summation-formula}
\end{eqnarray}
where $k$ is an integer number varying in the interval
\begin{equation}
	-\frac{q}{2}+\frac{\Delta\phi}{\Phi_{0}}\le k\le \frac{q}{2}+\frac{\Delta\phi}{\Phi_{0}}  \   .
\end{equation}
The substitution of \eqref{summation-formula} into \eqref{WFcs2}, allow us to integrate over $\chi$ with the help of the formula given below \cite{Grad}
\begin{equation}
	\int_{0}^{\infty}dxx^{\mu-1}e^{-vx}K_{\nu}(x)=\frac{\sqrt{\pi}2^{\nu}\Gamma(\mu-\nu)\Gamma(\mu+\nu)}{\Gamma(\mu+1/2)(v+1)^{\mu+\nu}}F\Bigg(\mu+\nu,\nu+\frac{1}{2};\mu+1/2;\frac{v-1}{v+1}\Bigg) \ ,
	\label{IntegralP}
\end{equation}
where $F(a,b;c,z)$ represents the hypergeometric function \cite{Grad}. Thus, the Wightman function due to the string, in closed form, is given by
\begin{eqnarray}
	W_{\rm{cs}}^{\rm{dS}}(x,x^{\prime})&=&\frac{1}{2(2\pi)^{\frac{D+1}{2}}a^{D-1}} \Biggr\{\sum_{k}e^{i\alpha_0(2\pi k -q\Delta\phi)}F_{\nu}^{D/2}(u_k)\nonumber\\%
	&-&\frac{e^{-iqn_{0}\Delta\phi}}{2\pi i}\sum_{j=\pm1}je^{ji\pi q|\alpha_0|}\int_{0}^{\infty}dy\frac{\cosh{[qy(1-|\alpha_0|)]}-\cosh{(|\alpha_0| qy)e^{-iq(\Delta\phi+j\pi)}}}{\cosh{(qy)}-\cos{(q(\Delta\phi+j\pi))}}\nonumber\\%
	&\times&F_{\nu}^{D/2}(u_y)\Biggl\} \ ,
	\label{dWFcs3}
\end{eqnarray}
where we have introduced the function
\begin{equation}
		F_{\nu}^{D/2}(v)=\frac{2^{\nu+1/2}\Gamma(D/2-\nu)\Gamma(D/2+\nu)}{\Gamma(D/2+1/2)(v+1)^{D/2+\nu}}F\Bigg(\frac{D}{2}+\nu,\nu+\frac{1}{2};\frac{D+1}{2};\frac{v-1}{v+1}\Bigg) \ ,
		\label{F-function}
\end{equation}
and the variables
\begin{eqnarray}
	u_k&=&-1+\frac{r^2+r^{\prime2}-2rr^{\prime}\cos(\Delta\phi+2\pi k/q)+\Delta\vec{x}_{||}^{2}+\Delta z^2-(\eta-\eta^{\prime})^2}{2\eta\eta^{\prime}} \ ,\nonumber\\%
	u_y&=&-1+\frac{r^2+r^{\prime2}+2rr^{\prime}\cosh(y/2)+\Delta\vec{x}_{||}^{2}+\Delta z^2-(\eta-\eta^{\prime})^2}{2\eta\eta^{\prime}} \ .
\end{eqnarray}

Now we turn to the boundary-induced term. Since the integral over $k_z$ unfortunately can be performed for an arbitrary value of the parameter $\beta$, we will study the particular cases of the Dirichlet and Neumann boundary conditions. According to \eqref{FBC}, for Dirichlet BC ($\beta=0$) we have $\alpha_{1}(k_z)=\pi/2$ and for Neumann BC ($\beta\rightarrow\infty$), $\alpha_{1}(k_z)=0$. Therefore, for these boundary conditions, the Wightman function due to the plate is
\begin{eqnarray}
	W_{\rm{pl}}(x,x^{\prime})&=&\delta_{(\rm{J})}\frac{q(\eta \eta^{\prime})^{D/2}}{2\pi^{\frac{D+3}{2}}a^{D-1}} \int_{0}^{\infty}d\chi\chi^{(D-3)/2}e^{-(r^2+r^{\prime2}+\Delta\vec{x}_{||}^{2}-\eta^2-\eta^{\prime2})\chi}K_{\nu}(2\eta\eta^{\prime}\chi)\nonumber\\%
	&\times&\sum_{n=-\infty}^{\infty}e^{inq\Delta\phi}I_{q|n+\alpha|}\left(2rr^{\prime}\chi\right)\int_{0}^{\infty}dk_{z}e^{- k_{z}^2/4\chi}\cos[k_{z}(z+z^{\prime})] \ ,
	\label{WFb2}
\end{eqnarray}
where $\rm{J}=D$ for Dirichlet BC, $\delta_{(\rm{J})}=1$, and $\rm{J}=N$ for Neumann BC, $\delta_{(\rm{J})}=-1$. Now integrating over $k_z$, using the formula \eqref{summation-formula} for the sum over $n$ and subsequently integrating over the variable $\chi$, we arrive to
\begin{eqnarray}
	W_{\rm{pl}}(x,x^{\prime})&=&\frac{\delta_{(\rm{J})}}{2(2\pi)^{\frac{D+1}{2}}a^{D-1}} \Biggr\{\sum_{k}e^{i\alpha_0(2\pi k -q\Delta\phi)}F_{\nu}^{D/2}(v_k)\nonumber\\%
	&-&\frac{e^{-iqn_{0}\Delta\phi}}{2\pi i}\sum_{j=\pm1}je^{ji\pi q|\alpha_0|}\int_{0}^{\infty}dy\frac{\cosh{[qy(1-|\alpha_0|)]}-\cosh{(|\alpha_0| qy)e^{-iq(\Delta\phi+j\pi)}}}{\cosh{(qy)}-\cos{(q(\Delta\phi+j\pi))}}\nonumber\\%
	&\times&F_{\nu}^{D/2}(v_y)\Biggl\} \ ,
	\label{WFb3}
\end{eqnarray}
with the arguments
\begin{eqnarray}
	v_k&=&-1+\frac{r^2+r^{\prime2}-2rr^{\prime}\cos(\Delta\phi+2\pi k/q)+\Delta\vec{x}_{||}^{2}+(z+z^{\prime})^2-(\eta-\eta^{\prime})^2}{2\eta\eta^{\prime}} \ ,\nonumber\\%
	v_y&=&-1+\frac{r^2+r^{\prime2}+2rr^{\prime}\cosh(y/2)+\Delta\vec{x}_{||}^{2}+(z+z^{\prime})^2-(\eta-\eta^{\prime})^2}{2\eta\eta^{\prime}} \ .
\end{eqnarray}
\section{Field squared}
\label{phi_2}
In this section we want to investigate the VEV of the field squared. This quantity is in the calculation of the energy-momentum tensor as we will see. Formally it is defined by evaluating the Wightman at the coincidence limit:
\begin{equation}
	\langle |\varphi|^2\rangle=\lim_{x\rightarrow x^{\prime}}W(x,x^{\prime}) \ ,
\end{equation}
where $|\varphi|^2$ is here understood as $\hat{\varphi}(x)\hat{\varphi}^{\dagger}(x)$. According to \eqref{WF_terms}, we decompose this VEV in the form
\begin{equation}
	\langle |\varphi|^2\rangle=\langle |\varphi|^2\rangle_{\rm{dS}}+\langle |\varphi|^2\rangle_{\rm{cs}}+\langle |\varphi|^2\rangle_{\rm{pl}} \ .
	\label{phi2-decomp}
\end{equation}
The contribution $\langle |\varphi|^2\rangle_{\rm{dS}}$ comes from the term $k=0$ in \eqref{dWFcs3}, and it is induced in the pure de Sitter spacetime. This term is divergent and independent of the spacetime coordinates. This contribution has been already investigated in \cite{Bunch1978, Candelas1975, Dowker1976a}. Specifically for $D=3$ this renormalized VEV is given by 
\begin{eqnarray}	\langle|\varphi|^2\rangle_{\rm{dS}}&=&\frac1{8\pi^2a^2}\left\{\left(\frac{m^2a^2}2+6\xi-1\right)\left[\Psi\left(\frac32+\nu\right)+\Psi\left(\frac32-\nu\right)
	-2\ln(m a)\right]\right.\nonumber\\
	&-&\left.\frac{(6\xi-1)^2}{m^2a^2}+\frac1{30m^2a^2}-6\xi+\frac23\right\} \ ,
\end{eqnarray}
where $\Psi(x)$ is the logarithimic derivative of the gamma-function. 

 The second contribution, $\langle|\varphi|^2\rangle_{\rm{cs}}$, is induced by the magnetic-flux-carrying cosmic string in the dS spacetime. Because the presence of the cosmic string does not introduce additional curvature, this contribution is finite for points with $r\neq 0$.\footnote{This contribution considering real bosonic field has been analyzed in Ref. \cite{Mello_09}.} Explicitly, the VEV of the field squared induced by the cosmic string, taking into account the interaction of the bosonic field with the magnetic flux, reads:
\begin{eqnarray}
	\langle |\varphi|^2\rangle_{\rm{cs}}&=&\frac{1}{(2\pi)^{\frac{D+1}{2}}a^{D-1}}\Biggr[\sideset{}{'}\sum_{k=1}^{[q/2]}\cos(2\pi k\alpha_0)F_{\nu}^{D/2}(2r_p^2s_k^2-1)\nonumber\\%
	&-&\frac{q}{2\pi}\int_{0}^{\infty}dy\frac{h(q,\alpha_0,y)}{\cosh{(qy)}-\cos{(q\pi)}}F_{\nu}^{D/2}(2r_p^2c_y^2-1)\Biggl] \ ,
	\label{FScs}
\end{eqnarray}
where $[q/2]$ represents the integer part of $q/2$, and the prime on the sign of the summation over $k$ means that for even values of $q$, the term $k=q/2$ should be
taken with the coefficient $1/2$. Moreover, we have defined the function
\begin{equation}
	h(q,\alpha_0,y)=\cosh[qy(1-|\alpha_0|)]\sin(q\pi|\alpha_0|)+\cosh(qy|\alpha_0|)\sin[q\pi(1-|\alpha_0|)] \ ,
\end{equation}

with
\begin{eqnarray}
	&&s_{k}=\sin(\pi k/q) \nonumber\\
	&&c_y=\cosh(y/2) \ .
	\label{sub-arg}
\end{eqnarray}
In \eqref{FScs} , $r_p=r/\eta$ and $z_p=z/\eta$ are the proper distances from the string and the plate, respectively, in unities of the dS spacetime curvature, $a$. 

As to the last term in the r.h.s. of \eqref{phi2-decomp}, it can be yet decomposed in two contributions:
\begin{equation}
	\langle |\varphi|^2\rangle_{\rm{pl}}=\langle |\varphi|^2\rangle_{\rm{pl}}^{(0)}+\langle |\varphi|^2\rangle_{\rm{pl}}^{(q,\alpha_0)} \ .
	\label{FS_pl_decomp}
\end{equation}
The first contribution on the r.h.s. of the above expression is the term induced purely by the plate in the dS spacetime and has been analyzed in \cite{Saharian:2009ii}. Explicitly, for Dirichlet or Neumann boundary conditions, it is given by
\begin{eqnarray}
	\langle |\varphi|^2\rangle_{\rm{pl}}^{(0)}&=&\frac{\delta_{(\rm{J})}}{(2\pi)^{\frac{D+1}{2}}a^{D-1}}F_{\nu}^{D/2}(2z_p^2-1) \ .
	\label{FS-pl}
\end{eqnarray}
The second contribution, induced by the magnetic-flux-carrying cosmic string and the plate, is a new one and is given by 
\begin{eqnarray}
	\langle |\varphi|^2\rangle_{\rm{pl}}^{(q,\alpha_0)}&=&\frac{\delta_{(\rm{J})}}{(2\pi)^{\frac{D+1}{2}}a^{D-1}}\Biggl[\sideset{}{'}\sum_{k=0}^{[q/2]}\cos(2\pi k\alpha_0)F_{\nu}^{D/2}(w_k)\nonumber\\%
	&-&\frac{q}{2\pi}\int_{0}^{\infty}dy\frac{h(q,\alpha_0,y)}{\cosh{(qy)}-\cos{(q\pi)}}F_{\nu}^{D/2}(w_y)\Biggr] \ ,
	\label{FS-cs-pl}
\end{eqnarray} 
where we have introduced the notations
\begin{eqnarray}
	w_k&=&2r_p^2s_k^2+2z_p^2-1 \nonumber\\%
	w_y&=&2r_p^2c_y^2+2z_p^2-1 \ .
	\label{arguments}
\end{eqnarray}

Let us now investigate the behavior of the field squared for some special and asymptotic cases. We would like to remind that in this paper, our goal is the study of the new contribution induced by the presence of the cosmic string and the plate. For a conformal massless quantum scalar field, we have $\nu=1/2$ according to \eqref{nu}, and in this case we have the relation
\begin{equation}
	F_{1/2}^{D/2}(x)=\frac{\Gamma(D/2-1/2)}{(1+x)^{(D-1)/2}} \ ,
	\label{F-massless}
\end{equation}
which replaced in \eqref{FS-cs-pl}, yields
\begin{eqnarray}
	\langle |\varphi|^2\rangle_{\rm{pl}}^{(q,\alpha_0)}&=&\delta_{(\rm{J})}\frac{2\Gamma(D/2-1/2)}{(4\pi)^{\frac{D+1}{2}}}\Big(\frac{\eta}{a}\Big)^{D-1}\Biggr[\sideset{}{'}\sum_{k=1}^{[q/2]}\frac{\cos(2\pi k\alpha_0)}{(r^2s_k^2+z^2)^{(D-1)/2}}\nonumber\\%
	&-&\frac{q}{2\pi}\int_{0}^{\infty}dy\frac{h(q,\alpha_0,y)}{[\cosh{(qy)}-\cos{(q\pi)}](r^2c_y^2+z^2)^{(D-1)/2}}\Biggl] \ .
	\label{FSCm0}
\end{eqnarray}
This result is conformally related to the VEV of field squared induced in the Minkowski bulk with a plate placed at $z=0$ and a cosmic string with a magnetic flux, for both Dirichlet and Neumann boundary conditions (see Ref. \cite{Braganca:2020jci}).

We now analyze the behaviour of $\langle|\varphi^2|\rangle_{\rm{pl}}^{(q,\alpha_0)}$ for points close and far from the string. For points outside of the plate, $z\neq0$, the VEV of the field squared on the string is finite and can be obtained directly by putting $r=0$ in \eqref{FS-cs-pl}. In the opposite limit, i.e., for points far from string we need the asymptotic expression for large arguments of the function $F_{\nu}^{D/2}(x)$. In the leading term, we have the approximate formula:
\begin{equation}
	F_{\nu}^{D/2}(x)\approx\frac{2^{\nu-1/2}}{\sqrt{\pi}}\Gamma(\nu)\Gamma(D/2-\nu)x^{-D/2+\nu} \ .
	\label{Asymp-F}
\end{equation}
Therefore, for distant points from the string, $r\gg \eta,z$, we can use the above asymptotic formula in \eqref{FS-cs-pl},  yielding
\begin{eqnarray}
	\langle |\varphi|^2\rangle_{\rm{pl}}^{(q,\alpha_0)}&\approx&\delta_{(\rm{J})}\frac{2^{2\nu-D-1}\Gamma(\nu)\Gamma(D/2-\nu)}{\pi^{\frac{D}{2}+1}a^{D-1}r_{p}^{D-2\nu}}\Biggr[\sideset{}{'}\sum_{k=1}^{[q/2]}\frac{\cos(2\pi k\alpha_0)}{s_k^{D-2\nu}}\nonumber\\%
	&-&\frac{q}{2\pi}\int_{0}^{\infty}dy\frac{c_y^{-(D-2\nu)}h(q,\alpha_0,y)}{\cosh{(qy)}-\cos{(q\pi)}}\Biggl] \ .
	\label{FSb3}
\end{eqnarray}

We now turn to the behaviour of the field squared induced by the string and plate, regarding the limiting regions along the $z$-axis. On the surface of the plate, $z=0$, $\langle |\varphi|^2\rangle_{\rm{pl}}^{(q,\alpha_0)}$ is finite for points outside the string, $r\neq0$. On the other hand, far from the plate location, $z\gg \eta,r$, we can once again make use of \eqref{Asymp-F}, obtaining
\begin{eqnarray}
	\langle |\varphi|^2\rangle_{\rm{pl}}^{(q,\alpha_0)}&\approx&\delta_{(\rm{J})}\frac{2^{2\nu-D-1}\Gamma(\nu)\Gamma(D/2-\nu)}{\pi^{\frac{D}{2}+1}a^{D-1}z_{p}^{D-2\nu}}\Biggr[\sideset{}{'}\sum_{k=1}^{[q/2]}\cos(2\pi k\alpha_0)\nonumber\\%
	&-&\frac{q}{2\pi}\int_{0}^{\infty}dy\frac{h(q,\alpha_0,y)}{\cosh{(qy)}-\cos{(q\pi)}}\Biggl] \ .
	\label{FSb4}
\end{eqnarray}

Finally, we consider the Minkowskian limit, $a\rightarrow\infty$, with fixed value of $t$. In this limit, the geometry under consideration is reduced to the geometry
of a cosmic string in the background of $(D+1)$-dimensional Minkowski spacetime. For the analysis of this limit, the representation for the field squared given in \eqref{FS-cs-pl} is not convenient. Thus, for this purpose, we return to the Wightman function representation induced by the plate given in \eqref{WFb2}. For the
coordinate $\tau$ in the arguments of the modified Bessel function one has $\eta\approx |t-a|$. Also in this limit we have $\nu\gg1$, which is imaginary and, according to \eqref{nu}, is $\nu\approx ima$. By making use of the uniform asymptotic expansion for the Macdonald function of imaginary order given in \cite{Balogh1967}, replacing it in \eqref{WFb2} and subsequently taking the coincidence limit, after some intermediate steps, we get
\begin{eqnarray}
	\langle |\varphi|^2\rangle_{\rm{pl}}^{(q,\alpha_0),(\rm{M})}&=&\delta_{(\rm{J})}\frac{2m^{D-1}}{(2\pi)^{\frac{D+1}{2}}}\Biggl[\sideset{}{'}\sum_{k=1}^{[q/2]}\cos(2\pi k\alpha_0)f_{\frac{D-1}{2}}\Big(2m\sqrt{r^2s_k^2+z^2}\Big)\nonumber\\%
	&-&\frac{q}{2\pi}\int_{0}^{\infty}dy\frac{h(q,\alpha_0,y)}{\cosh{(qy)}-\cos{(q\pi)}}f_{\frac{D-1}{2}}\Big(2m\sqrt{r^2c_y^2+z^2}\Big)\Biggr] \ ,
	\label{FS-M}
\end{eqnarray}
with the notation
\begin{equation}
	f_{\nu}(x)=\frac{K_{\nu}(x)}{x^{\nu}} \ .
	\label{f_function}
\end{equation}
Note that this result exactly coincides with the corresponding VEV of the field squared calculated in the $(D+1)$-dimensional Minkowskian bulk \cite{Braganca:2020jci}.

\begin{figure}[!htb]
	\begin{center}
		\includegraphics[scale=0.4]{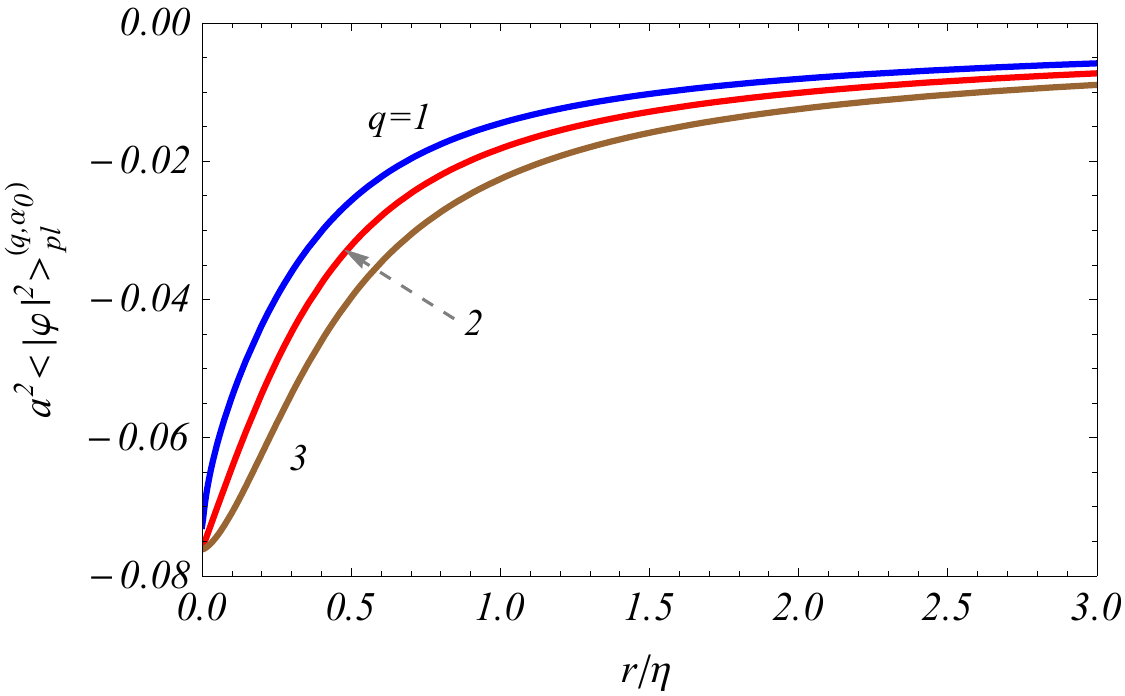}
		\quad
		\includegraphics[scale=0.4]{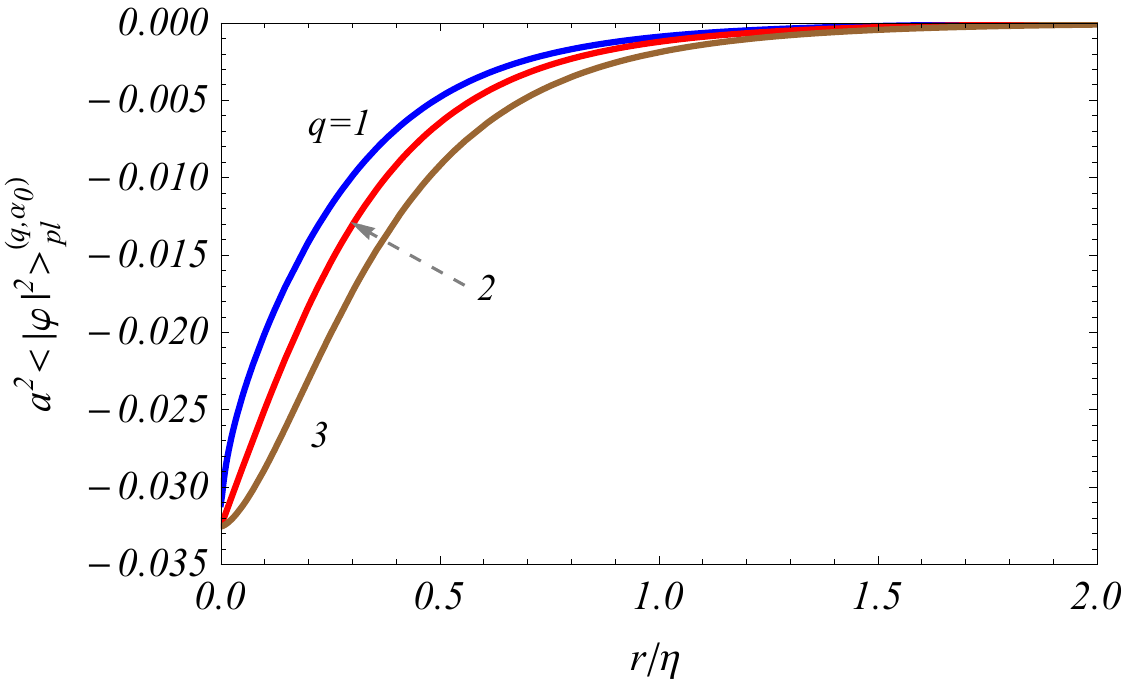}
		\caption{The VEV of the complex scalar field squared induced by the plate for $D=3$, Eq. \eqref{FS-cs-pl}, is plotted in terms of the proper distance from the cosmic string in units of $a$, $r/\eta$, for $z/\eta=0.5$ $\alpha_0=0.25$, $\xi=0$ and $q=1$, 2 and 3. The plot on the left is for $ma=1$, while the plot on the right is for $ma=2$. The numbers near the curves are the values of $q$.}
		\label{fig1}
	\end{center}
\end{figure}

In Fig.\ref{fig1} we plot the behavior the field quared induced by the plate, $\langle |\varphi|^2\rangle_{\rm{pl}} ^{(q,\alpha_0)}$, as function of $r/\eta$ for $D=3$, considering minimum coupling, $\xi=0$, and fixed parameters $z/\eta=0.5$ and $\alpha_0=0.25$, for different values of the parameter $q$. In the left panel we adopt $ma=1$ (real $\nu$) and in the right panel $ma=2$ (imaginary $\nu$). We note from these graphs that the VEV is finite on the cosmic string and goes to zero far from it. From these plots we also note that the effect of the deficit angle due to the string is to enhance the intensity of the VEV of the field squared. Moreover, comparison between the panel shows that the intensity decreases with the mass.

\begin{figure}[!htb]
	\begin{center}
		\includegraphics[scale=0.4]{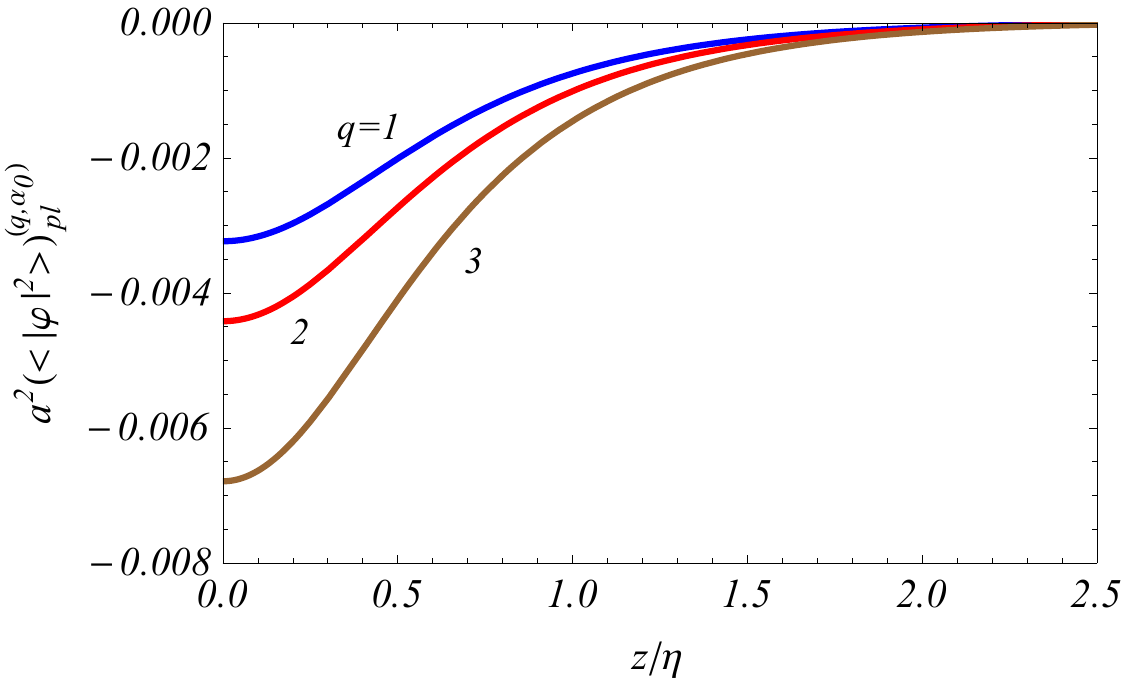}
		\quad
		\includegraphics[scale=0.4]{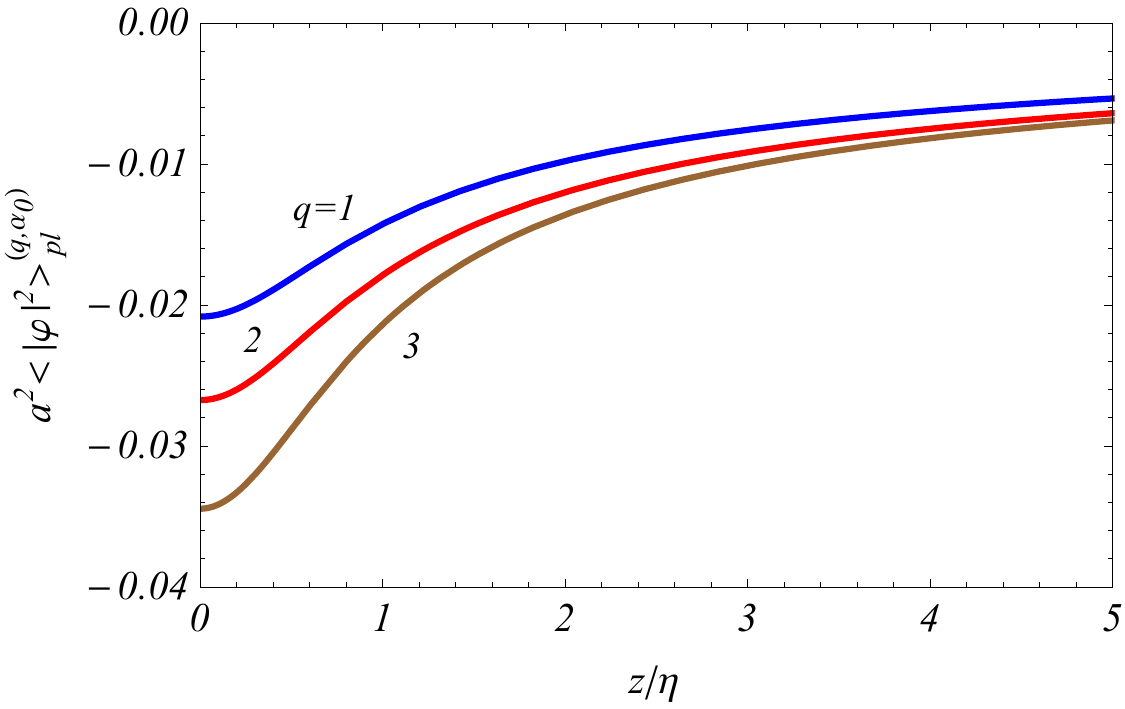}
		\caption{The VEV of the field squared induced by the plate, Eq. \eqref{FS-cs-pl}, is plotted for $D=3$ as function of $z/\eta$, for fixed $\xi=0$, $\alpha_0=0.25$ and $r/\eta=0.75$. The numbers near the curves correspond to the values of deficit angle parameter, $q$. The left panel corresponds to $ma=1$ and the right panel to $ma=2$. Numbers near the curves are the values of $q$.}
		\label{fig2}
	\end{center}
\end{figure}

In Fig. \ref{fig2}, we present the behavior of \eqref{FS-cs-pl} as function of proper distance from the plate in units of $a$, $z/\eta$, for $D=3$ and fixed values $r/\eta=0.75$, $\alpha_0=0.25$ and $\xi=0$, considering different values of $q$. In the left panel we assume $ma=1$, and in the right panel $ma=2$. These plots shows that the VEV of the field squared is finite on the plate and vanishes far from it.
\section{Energy-momentum tensor}
\label{EMT}
We now turn to the calculation of the VEV of the energy-momentum tensor in the setup under consideration. For this end, we use the formula \cite{Wagner_20}
\begin{equation}
	\langle T_{\mu\nu}\rangle=\lim_{x^{\prime}\rightarrow x}(D_{\mu}D_{\nu'}^{\dagger}+D_{\mu'}^{\dagger}D_{\nu})W(x,x')-2[\xi R_{\mu\nu}+\xi\nabla_{\mu}\nabla_{\nu}-(\xi-1/4)g_{\mu\nu}\nabla_{\alpha}\nabla^{\alpha}]]\langle|\varphi|^2\rangle \ ,
	\label{E-M-Tensor-formula}
\end{equation}
where $R_{\mu\nu}=Dg_{\mu\nu}/a^2$ is the Ricci tensor for the dS spacetime and $D_{\mu}=\nabla_{\mu}+ieA_{\mu}$. Similarly to the VEV of the field squared, the VEV of the energy-momentum tensor can be decomposed as
\begin{equation}
	\langle T_{\mu\nu}\rangle=\langle T_{\mu\nu}\rangle_{\rm{dS}}+\langle T_{\mu\nu}\rangle_{\rm{cs}}+\langle T_{\mu\nu}\rangle_{\rm{pl}} \  .
	\label{TEM0}
\end{equation}
The first contribution in the r.h.s. of the above equation, $\langle T_{\mu\nu}\rangle_{\rm{dS}}$, is induced in the pure dS spacetime and is a coordinate independent quantity, due to the maximal symmetry of dS spacetime and the vacuum state under consideration. Additionally, this contribution is divergent and some renormalization scheme is needed to provide a finite result. This procedure has been considered in \cite{Bunch1978, Candelas1975, Dowker1976a}. The second term on the r.h.s., $\langle T_{\mu\nu}\rangle_{\rm{cs}}$, is induced by the cosmic string and the magnetic flux along its core and it is finite away from the string.\footnote{For the case of real scalar field, this VEV has been calculated in \cite{Mello_09}.}  Here we decided do not exhibit this contribution in order do not extend this paper too long, since our propose is to investigate the energy-momentum tensor induced by the plate only, $\langle T_{\mu\nu}\rangle_{\rm{pl}}$. The latter can be decomposed in two contributions:
\begin{equation}
	\langle T_{\mu\nu}\rangle_{\rm{pl}}=\langle T_{\mu\nu}\rangle_{\rm{pl}}^{(0)}+\langle T_{\mu\nu}\rangle_{\rm{pl}}^{(q,\alpha_0)} \ ,
\end{equation}
where the first contribution on the r.h.s. is induced purely by the plate in dS spacetime and has been analyzed in \cite{Saharian:2009ii}. The second term is a new contribution and is induced by the string and the plate. This contribution is finite away from the string or from the plate location, as will see below. In what follows our analysis is constrained to the latter contribution.

The covariant d'Alembertian acting in \eqref{FS-cs-pl} provides
\begin{eqnarray}
	\Box\langle|\varphi|^2\rangle_{\rm{pl}}^{(q,\alpha_0)}&=&-\frac{4\delta_{(\rm{J})}}{(2\pi)^{\frac{D+1}{2}}a^{D+1}}\Biggl\{\sideset{}{'}\sum_{k=1}^{[q/2]}\cos(2\pi k/q)\Bigg[\Big[1+2\gamma^2-(D+2)(s_k^2r_p^2+z_p^2)\Big]\nonumber\\
	&\times&\frac{\partial}{\partial y}F_{\nu}^{D/2}(y)+4\Big[(s_k^4r_p^2+z_p^2)-(s_k^2r_p^2+z_p^2)^2\Big]\frac{\partial^2}{\partial y^2}F_{\nu}^{D/2}(y)\Bigg]\Bigg|_{y=w_{k}}\nonumber\\
	&-&\frac{q}{2\pi}\int_{0}^{\infty}dy\frac{h(q,\alpha_0,y)
	}{\cosh{(qy)}-\cos{(q\pi)}}\Bigg[\Big[1+2\gamma^2-(D+2)(c_y^2r_p^2+z_p^2)\Big]\nonumber\\
	&\times&\frac{\partial}{\partial y}F_{\nu}^{D/2}(y)+4\Big[(c_y^4r_p^2+z_p^2)-(c_y^2r_p^2+z_p^2)^2\Big]\frac{\partial^2}{\partial y^2}F_{\nu}^{D/2}(y)\Bigg]\Bigg|_{y=w_{y}}\Biggr\} \ ,
\end{eqnarray}
with $w_k$ and $w_y$ been given by \eqref{arguments}.

For the geometry under consideration, only the $\nabla_{\eta}\nabla_{r}$, $\nabla_{r}\nabla_{z}$, $\nabla_{\eta}\nabla_{z}$ and $\nabla_{\mu}\nabla_{\mu}$ differential operators contribute when acting on the VEV of the field squared.
The remaining contributions come from the electromagnetic covariant derivatives acting on the Wightman function. As to the azimuthal component, it is more convenient to act the $D_{\phi}D_{\phi'}^{\dagger}$ operator in \eqref{WFb2}, and subsequently take the coincidence limit in all coordinates, including the angular one. Following this procedure, we obtain:
\begin{equation}
	\mathcal{I}(q,\alpha,y)=\sum_{n=-\infty}^{\infty}q^2(n+\alpha)^2I_{q|n+\alpha|}(y) \ ,
\end{equation}
where $y=2rr'\chi$. This sum can be developed by using the differential equation obeyed by the modified Bessel function. Then we get,
\begin{equation}
	\mathcal{I}(q,\alpha,y)=\bigg(y^2\frac{d^2}{dy^2}+y\frac{d}{dy}-y^2\bigg)\sum_{n=-\infty}^{\infty}I_{q|n+\alpha|}(y) \ ,
\end{equation}
where this last sum is given by \cite{Braganca:2014qma}
\begin{equation}
	\sum_{n=-\infty}^{\infty}I_{q|n+\alpha|}(y)=\frac{e^{y}}{q}+\frac{2}{q}\sideset{}{'}\sum_{k=1}^{[q/2]}\cos(2\pi k\alpha_0)e^{y\cos(2\pi k/q)}-\frac{1}{\pi}\int_{0}^{\infty}dy\frac{e^{-y\cosh(y)}f(q,\alpha_0,y)}{\cosh(qy)-\cos(q\pi)} \ .
\end{equation}
Note that the first term on the r.h.s. is induced purely by the plate in dS spacetime and accounts for the contribution $\langle T_{\nu}^{\mu}\rangle_{\rm{pl}}^{(0)}$ only and, therefore, is not taken into account in further analysis. 

The boundary induced contribution in the VEV of the energy-momentum tensor is calculated by making use of the corresponding parts in the Wightman function and VEV of the field squared. After long but straightforward steps, we get (no summation over $\mu$)
\begin{eqnarray}
	\langle T_{\mu}^{\mu}\rangle_{\rm{pl}}^{(q,\alpha_0)}&=&-\frac{2\delta_{(\rm{J})}}{(2\pi)^{\frac{D+1}{2}}a^{D+1}}\Bigg[\sideset{}{'}\sum_{k=1}^{[q/2]}\cos(2\pi k\alpha_0)G_{\mu}^{\mu}(s_k,r_p,z_p,w_k)\nonumber\\
	&-&\frac{q}{2\pi}\int_{0}^{\infty}dy\frac{h(q,\alpha_0,y)}{\cosh(qy)-\cos(q\pi)}G_{\mu}^{\mu}(c_y,r_p,z_p,w_y)\Bigg]  \  ,
	\label{EM-Plate}
\end{eqnarray}
with the functions
\begin{eqnarray}
	G_{0}^{0}(\gamma,u,v,y)&=&D\xi F_{\nu}^{D/2}(y)+\Big[-1+\xi_1(1+2\gamma^2)-D\xi_1	(\gamma^2u^2+v^2)\Big]\nonumber\\
	&\times&\frac{\partial}{\partial y}F_{\nu}^{D/2}(y)+4\xi_1(\gamma^4u^2+v^2)\frac{\partial^2}{\partial y^2}F_{\nu}^{D/2}(y) \ , \nonumber\\
	G_{1}^{1}(\gamma,u,v,y)&=&D\xi F_{\nu}^{D/2}(y)+\Bigl\{-2+4\xi(1+\gamma^2)+\big[1-(D+1)\xi_1\big](\gamma^2u^2+v^2)\Bigr\}\nonumber\\
	&\times&\frac{\partial}{\partial y}F_{\nu}^{D/2}(y)+4\xi_1\Big[v^2-(\gamma^2u^2+z^2)^2\Big]\frac{\partial^2}{\partial y^2}F_{\nu}^{D/2}(y) \ ,
	\nonumber\\
	G_{2}^{2}(\gamma,u,v,y)&=&D\xi F_{\nu}^{D/2}(y)+\Bigl\{-1+\gamma^2+\xi_1(1+\gamma^2)+\big[1-(D+1)\xi_1\big](\gamma^2u^2+v^2)\Bigr\}\nonumber\\
	&\times&\frac{\partial}{\partial y}F_{\nu}^{D/2}(y)+
	4\left\{(4	\xi\gamma^2-1)\gamma^2u^2+\xi_1\Big[v^2-(\gamma^2u^2+z^2)^2\Big]\right\}\frac{\partial^2}{\partial y^2}F_{\nu}^{D/2}(y) \ ,\nonumber\\
	G_{3}^{3}(\gamma,u,v,y)&=&D\xi F_{\nu}^{D/2}(y)+\Bigl\{2\xi_1\gamma^2+\big[1-(D+1)\xi_1\big](\gamma^2u^2+v^2)\Bigr\}\nonumber\\
	&\times&\frac{\partial}{\partial y}F_{\nu}^{D/2}(y)+4\xi_1\Big[\gamma^4u^2-(\gamma^2u^2+z^2)^2\Big]\frac{\partial^2}{\partial y^2}F_{\nu}^{D/2}(y) \ ,
	\nonumber\\
	G_{i}^{i}(\gamma,u,v,y)&=&D\xi F_{\nu}^{D/2}(y)+\Bigl\{-1+\xi_1(1+2\gamma^2)+\big[1-(D+1)\xi_1\big](\gamma^2u^2+v^2)\Bigr\}\nonumber\\
	&\times&\frac{\partial}{\partial y}F_{\nu}^{D/2}(y)+4\xi_1\Big[(\gamma^4u^2+v^2)-(\gamma^2u^2+z^2)^2\Big]\frac{\partial^2}{\partial y^2}F_{\nu}^{D/2}(y) \ ,
	\label{G-functions}
\end{eqnarray}
with $\xi_1=4\xi-1$. In the notation of the equation above, the indexes 0, 1, 2, 3 correspond to the coordinates $\eta$, $r$, $\phi$, $z$ and $i=4,...,D$ correspond to the extra dimensions ones. Additionally, we have three off-diagonal components,
\begin{eqnarray}
	G_{0}^{1}(\gamma,u,v,y)&=&u\gamma^2\Bigg[(\xi_1-1)\frac{\partial}{\partial y}F_{\nu}^{D/2}(y)+4\xi_1(u^2\gamma^2+v^2)\frac{\partial^2}{\partial y^2}F_{\nu}^{D/2}(y)\Bigg] \ ,
	\label{Off-diagonal01} \\
		G_{1}^{3}(\gamma,u,v,y)&=&-4\xi_1uv\gamma^2\frac{\partial^2}{\partial y^2}F_{\nu}^{D/2}(y)  \ ,
	\label{Off-diagonal2}\\
	G_{0}^{3}(\gamma,u,v,y)&=&v\Bigg[(\xi_1-1)\frac{\partial}{\partial y}F_{\nu}^{D/2}(y)+4\xi_1(u^2\gamma^2+v^2)\frac{\partial^2}{\partial y^2}F_{\nu}^{D/2}(y)\Bigg] \ .
	\label{Off-diagonal03}
\end{eqnarray}

Note that in \eqref{EM-Plate} the term $k=0$ is not included. As explained in the previous section, this is the boundary-induced term in the absence of the string and magnetic flux, which has been investigated in \cite{Saharian:2009ii}. Therefore, we are not considering the analysis of that term.

At this point we want to say that the above VEV of the energy-momentum tensor satisfies the trace relation:
\begin{equation}
	\label{trace}
	\langle T_{\mu}^{\mu}\rangle=2[D(\xi-\xi_c)\nabla_{\mu}\nabla^{\mu}\langle|\varphi|^2\rangle+m^2\langle|\varphi|^2\rangle] \ .
\end{equation}
In particular, this is zero for a conformal massless quantum scalar field. Moreover, due to the off-diagonal components, from the conservation condition, $\nabla_{\mu} \langle T_{\nu}^{\mu}\rangle=0$, three non-trivial differential equations take place:
\begin{eqnarray}
	\label{conservation}
	&&\partial_{\eta}\langle T_0^0\rangle+\frac{1}{\eta}[\langle T_{\mu}^{\mu}\rangle-(D+1)\langle T_0^0\rangle]+\frac{1}{r}\partial_{r}(r\langle T_0^1\rangle)+\partial_{z}\langle T_0^3\rangle=0 \ ,\nonumber\\
	&&\frac{1}{r}\partial_{r}(r\langle T_1^1\rangle)+\partial_{\eta}\langle T_1^0\rangle+\partial_{z}\langle T_1^3\rangle-\frac{1}{r}\langle T_2^2\rangle-\frac{(D+1)}{\eta}\langle T_1^0\rangle=0 \ , \nonumber\\
	&&\partial_{z}\langle T_3^3\rangle+\frac{1}{r}\partial_{r}(r\langle T_3^1\rangle)+\partial_{\eta}\langle T_3^0\rangle-\frac{(D+1)}{\eta}\langle T_3^0\rangle=0 \ . \nonumber\\
\end{eqnarray}
It can also be checked that these relations are satisfied by the components of the  VEV of the energy-momentum tensor above.

Let us now investigate some special and asymptotic cases of the energy density. We start analysing the conformal massless scalar field case. In this situation we have $\nu=1/2$. Using the corresponding formula for the function $F_{1/2}^{D/2}(x)$ given in \eqref{F-massless}, after some intermediate steps, we have:
\begin{eqnarray}
	\langle T_{0}^{0}\rangle_{\rm{pl}}^{(q,\alpha_0)}&=&-\delta_{(\rm{J})}\frac{2\Gamma(D/2+1/2)}{(4\pi)^{\frac{D+1}{2}}D}\left(\frac{\eta}{a}\right)^{D+1}\Bigg[\sideset{}{'}\sum_{k=1}^{[q/2]}\cos(2\pi k\alpha_0)s_k^2\frac{[2-(D-1)(s_k^2-1)]r^2+2z^2}{(r^2s_k^2+z^2)^{(D+3)/2}}\nonumber\\
	&-&\frac{q}{2\pi}\int_{0}^{\infty}dy\frac{c_y^2h(q,\alpha_0,y)}{\cosh(qy)-\cos(q\pi)}\frac{[2-(D-1)(c_y^2-1)]r^2+2z^2}{(r^2c_y^2+z^2)^{(D+3)/2}}\Bigg]  \  .
	\label{ED-Brane-massless}
\end{eqnarray}

We turn now to the study of the asymptotic cases. Near the string, $r\ll\eta,z$, for $z\neq0$ and $q|\alpha_0|>1$, the energy density is finite and we can directly put $r=0$. On the other hand, for $q|\alpha_0|<1$ the energy density on the string diverges as $\langle T_{0}^{0}\rangle_{\rm{pl}}^{(q,\alpha_0)}\sim 1/r_p^{2(1-q|\alpha_0|)}$.
Moreover, far from the string, $r\gg\eta,z$, we proceed in the same way as we did with the VEV of the field squared. Thus, to the leading order term, we get
\begin{eqnarray}
	\langle T_{0}^{0}\rangle_{\rm{pl}}^{(q,\alpha_0)}&\approx&-\delta_{(\rm{J})}\frac{2^{2\nu-D-2}D[1-\xi_1(D-2\nu-1)]\Gamma(\nu)\Gamma(D/2-\nu)}{\pi^{\frac{D}{2}+1}a^{D+1}r_p^{D-2\nu}}\nonumber\\
	&\times&\Bigg[\sideset{}{'}\sum_{k=1}^{[q/2]}\frac{\cos(2\pi k\alpha_0)}{s_k^{D-2\nu}}-\frac{q}{2\pi}\int_{0}^{\infty}dy\frac{c_y^{-(D-2\nu)}h(q,\alpha_0,y)}{\cosh(qy)-\cos(q\pi)}\Bigg]  \  .
	\label{ED-Asymp-r}
\end{eqnarray}
On the plate, $z=0$, the VEV of the energy density induced by the string and plate is finite for points outside, $r\neq0$. In the opposite limit, $z\gg\eta,r$, the corresponding asymptotic expression, to the leading order, is
\begin{eqnarray}
	\langle T_{0}^{0}\rangle_{\rm{pl}}^{(q,\alpha_0)}&\approx&-\delta_{(\rm{J})}\frac{2^{2\nu-D-2}D[1-\xi_1(D-2\nu-1)]\Gamma(\nu)\Gamma(D/2-\nu)}{\pi^{\frac{D}{2}+1}a^{D+1}z_p^{D-2\nu}}\nonumber\\
	&\times&\Bigg[\sideset{}{'}\sum_{k=1}^{[q/2]}\cos(2\pi k\alpha_0)-\frac{q}{2\pi}\int_{0}^{\infty}dy\frac{h(q,\alpha_0,y)}{\cosh(qy)-\cos(q\pi)}\Bigg]  \  .
	\label{ED-Asymp-z}
\end{eqnarray}

In the Minkowskian limit, $a\rightarrow\infty$, we follow the same procedure described in the previous section. Therefore, in this limit the energy density is given by:
\begin{eqnarray}
	\langle T_{0}^{0}\rangle_{\rm{pl}}^{(q,\alpha_0),(\rm{M})}&=&\delta_{(\rm{J})}\frac{2m^{D-1}}{(2\pi)^{\frac{D+1}{2}}}\Biggl\{\sum_{k=1}^{[q/2]}\cos(2\pi k\alpha_0)\Biggl[\Big(1-\xi_1(1+2\gamma^2)\Big)f_{\frac{D+1}{2}}\left(2m\sqrt{s_k^2r^2+z^2}\right)\nonumber\\
	&+&4\xi_1f_{\frac{D+3}{2}}\left(2m\sqrt{s_k^2r^2+z^2}\right)\Biggr]-\frac{q}{2\pi}\int_{0}^{\infty}dy\frac{h(q,\alpha_0,y)}{\cosh{(qy)}-\cos{(q\pi)}}\Biggl[\Big(1-\xi_1(1+2\gamma^2)\Big)\nonumber\\
	&\times&f_{\frac{D+1}{2}}\left(2m\sqrt{c_y^2r^2+z^2}\right)+4\xi_1f_{\frac{D+3}{2}}\left(2m\sqrt{c_y^2r^2+z^2}\right)\Biggr]\Biggr\} \ ,
	\label{ED-M}
\end{eqnarray}
with the function $f_{\nu}(x)$ defined in \eqref{f_function}. Note that the expression above exactly coincides with the VEV calculated in $(D+1)$-dimensional Minkowkian bulk \cite{Braganca:2020jci}.

\begin{figure}[!htb]
	\begin{center}
		\includegraphics[scale=0.4]{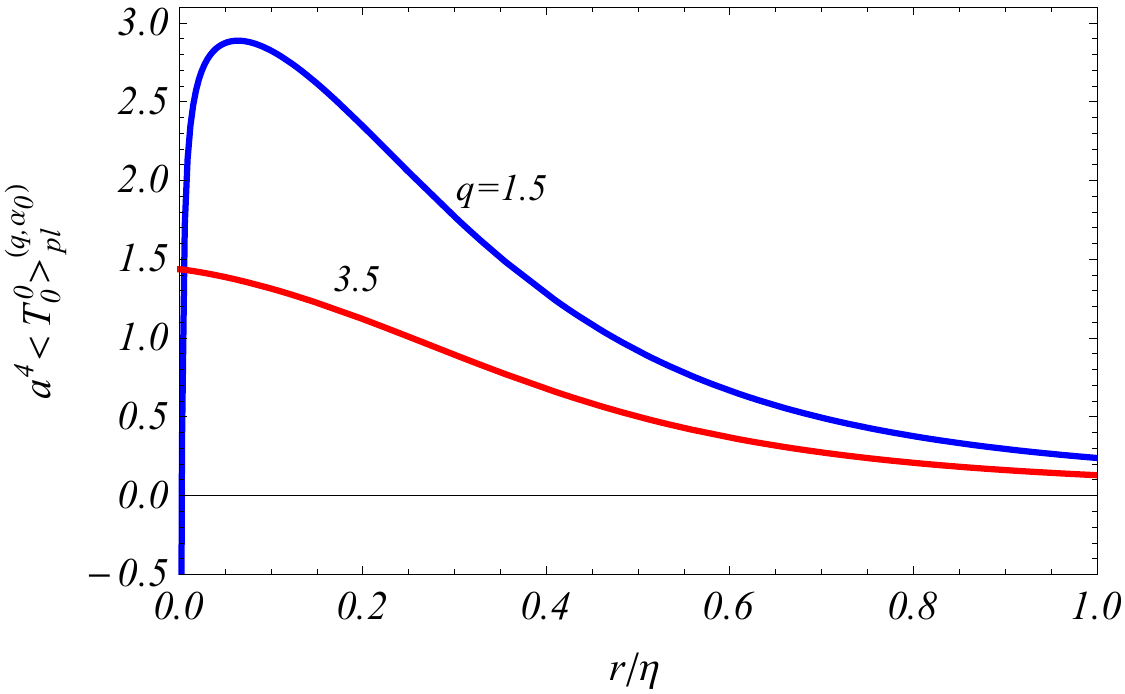}
		\quad
		\includegraphics[scale=0.4]{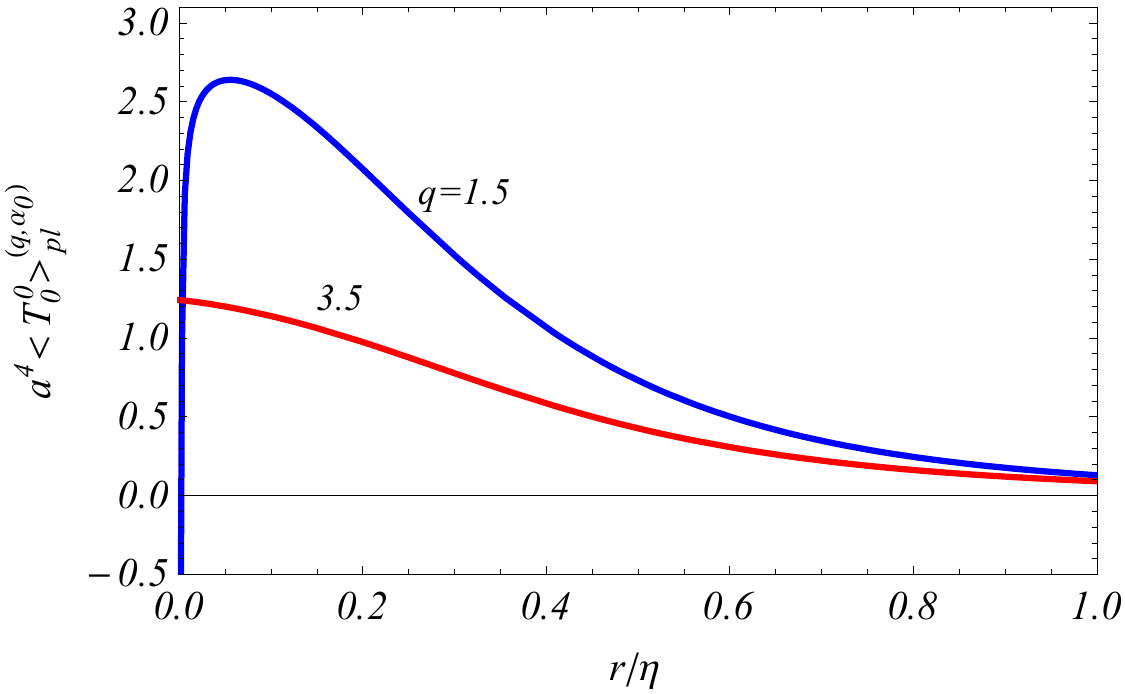}
		\caption{The energy density induced by the plate for $D=3$ is plotted in terms of the proper distance from the cosmic string, $r/\eta$, for $z/\eta=0.5$ $\alpha_0=0.45$, $\xi=0$ and $q=1$, 2 and 3. The plot on the left is for $ma=1$, while the plot on the right is for $ma=2$. The numbers near the curves are the values of $q$.}
		\label{fig3}
	\end{center}
\end{figure}

 In Fig. \ref{fig3} we exhibit the behavior the energy density induced by the plate as function of $r/\eta$ for $D=3$, considering minimum coupling, $\xi=0$, and fixed parameters $z/\eta=0.5$ and $\alpha_0=0.45$, for different values of the parameter $q$. In the left panel we adopt $ma=1$ (real $\nu$) and in the right panel $ma=2$ (imaginary $\nu$). As we can see from these plots, the energy density is divergent on the string for $q|\alpha_0|<1$ (curve with $q=1.5$) and is finite for $q|\alpha_0|>1$ (curve with $q=3.5$). This is in accordance to our asymptotic analysis for points near the string. Moreover, similarly to the field squared, the comparison between the plots shows that the intensity of the energy density decreases with the mass.

\begin{figure}[!htb]
	\begin{center}
		\includegraphics[scale=0.4]{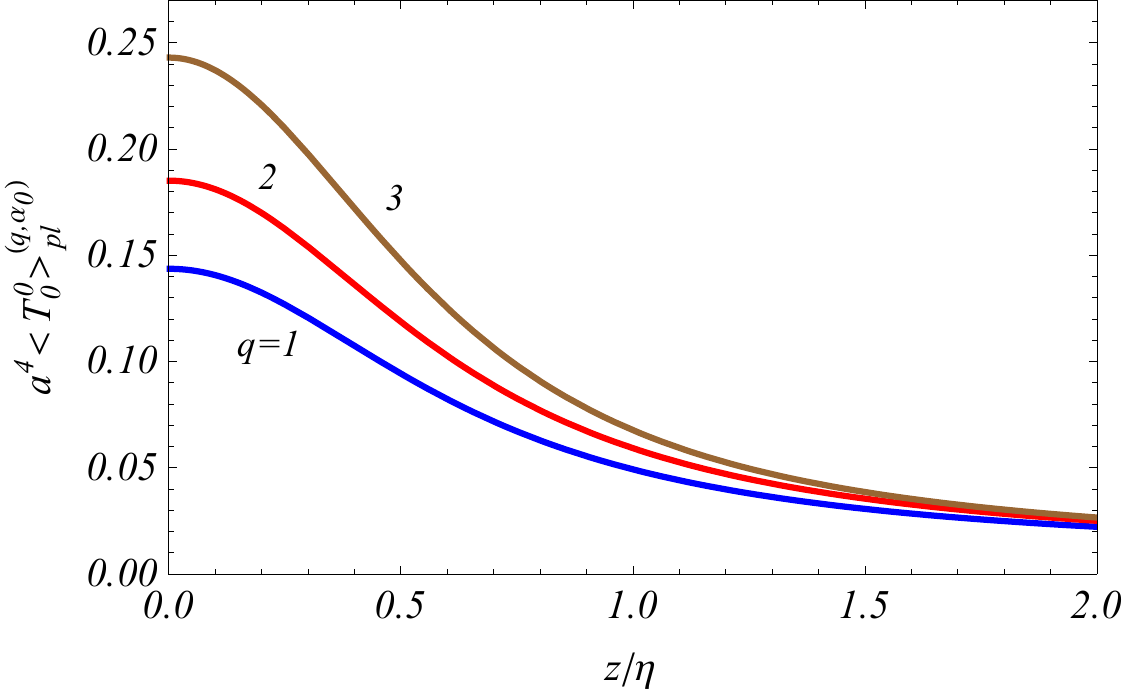}
		\quad
		\includegraphics[scale=0.4]{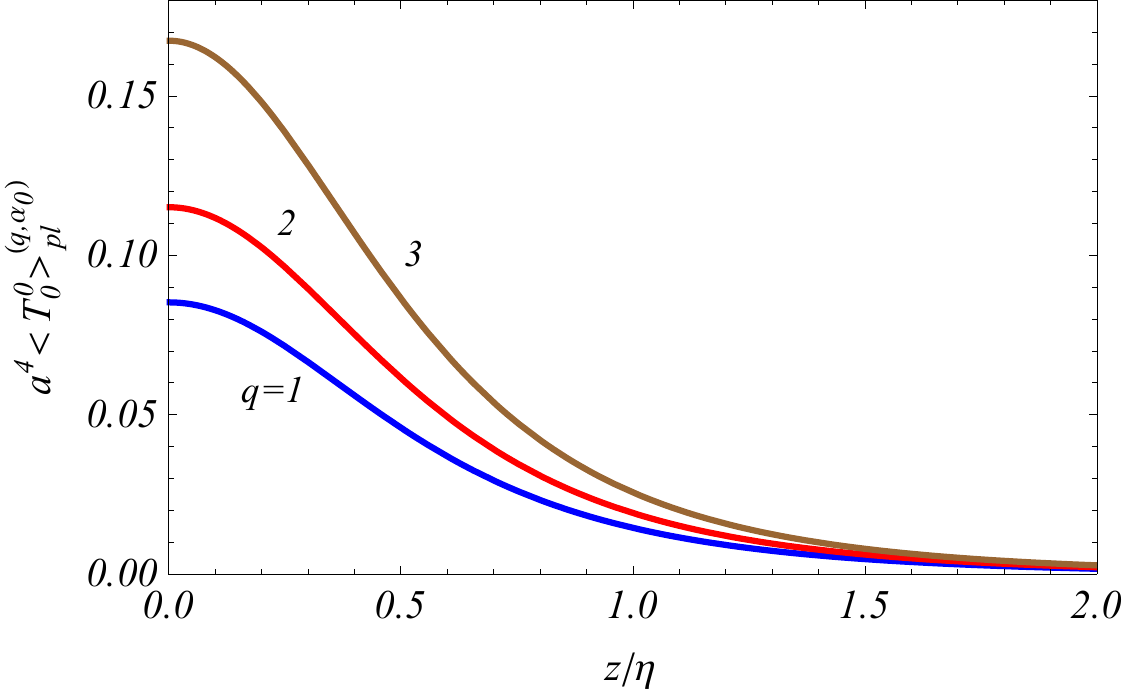}
		\caption{The VEV of the energy density induced by the plate is plotted for $D=3$ as function of $z/\eta$, for fixed $\xi=0$, $\alpha_0=0.25$ and $r/\eta=0.75$. The numbers near the curves correspond to the values of deficit angle parameter, $q$. The left panel corresponds to $ma=1$ and the right panel to $ma=2$. Numbers near the curves are the values of $q$.}
		\label{fig4}
	\end{center}
\end{figure}

In Fig. \ref{fig4} we present the behavior of $\langle T_{0}^{0}\rangle_{\rm{pl}}^{(q,\alpha_0)}$ as function of proper distance from the plate in units of $a$, $z/\eta$, for $D=3$ fixed values $r/\eta=0.75$, $\alpha_0=0.25$ and $\xi=0$, considering different values of $q$. In the left panel we assume $ma=1$, and in the right panel $ma=2$. These plots shows that the VEV of the energy density is finite on the plate and vanishes far from it.

The normal vacuum force acting on the plate is another interesting quantity present in the energy-momentum tensor given by \eqref{EM-Plate}. This quantity is given by the component $\langle T_{3}^{3}\rangle_{\rm{pl}}^{(q,\alpha_0)}$ at $z=0$ and is finite for $r\neq0$:
\begin{eqnarray}
	\langle T_{3}^{3}\rangle_{\rm{pl}}^{(q,\alpha_0)}\Big|_{z=0}&=&-\frac{2\delta_{(\rm{J})}}{(2\pi)^{\frac{D+1}{2}}a^{D+1}}\Bigg[\sideset{}{'}\sum_{k=1}^{[q/2]}\cos(2\pi k\alpha_0)G_{z}(s_k,2s_k^2r_p^2-1)\nonumber\\
	&-&\frac{q}{2\pi}\int_{0}^{\infty}dy\frac{h(q,\alpha_0,y)}{\cosh(qy)-\cos(q\pi)}G_{z}(c_y,2c_y^2r_p^2-1)\Bigg]  \  ,
	\label{EM33}
\end{eqnarray}
with
\begin{eqnarray}
	G_z(u,\gamma,y)&=&D\xi F_{\nu}^{D/2}(y)+\Bigl\{2\xi_1\gamma^2+\big[1-(D+1)\xi_1\big]\gamma^2u^2\Bigr\}\nonumber\\
	&\times&\frac{\partial}{\partial y}F_{\nu}^{D/2}(y)+4\xi_1\gamma^4u^2(1-u^2)\frac{\partial^2}{\partial y^2}F_{\nu}^{D/2}(y) \ .
\end{eqnarray}
We see that even in the case of $q=1$, there is a pressure on the flat plate caused by the integral term, which is a consequence of the interaction between the field and the magnetic flux running along the string's core. 

For a conformally coupled massless field, the vacuum effective pressure on the plate, $\langle T_{3}^{3}\rangle_{\rm{pl}}^{(q,\alpha_0)}\Big|_{z=0}$, can be even further simplified:

\begin{eqnarray}
	\label{presure_mass}
	\langle T_{3}^{3}\rangle_{\rm{pl}}^{(q,\alpha_0)}\Big|_{z=0}&=&\delta_{(\rm{J})}\frac{(D-1)^2}{(4\pi)^{\frac{D+1}{2}}D}\left(\frac{\eta}{ar}\right)^{D+1}\Bigg[\sideset{}{'}\sum_{k=1}^{[q/2]}\frac{\cos(2\pi k\alpha_0)}{s_k^{D-1}}\nonumber\\
	&-&\frac{q}{2\pi}\int_{0}^{\infty}dy\frac{c_y^{-(D-1)}h(q,\alpha_0,y)}{\cosh(qy)-\cos(q\pi)}\Bigg] \  .
\end{eqnarray}
Additionally, in the absence of magnetic flux, $\alpha_0=0$, and integer $q$, the second term on the r.h.s. of the above equation is absent, and for $D=3$, we have a simple expression,
\begin{equation}
	\label{pressure_0}
	\langle T_{3}^{3}\rangle_{\rm{pl}}^{(q,\alpha_0)}\Big|_{z=0}=\frac{\delta_{(\rm{J})}}{36\pi^{2}}\left(\frac{\eta}{ar}\right)^{4}(q^2-1) \ ,
\end{equation}
which is conformally related with the one calculated in the Minkowskian bulk \cite{Mello_11}.

\begin{figure}[!htb]
	\begin{center}
		\includegraphics[scale=0.4]{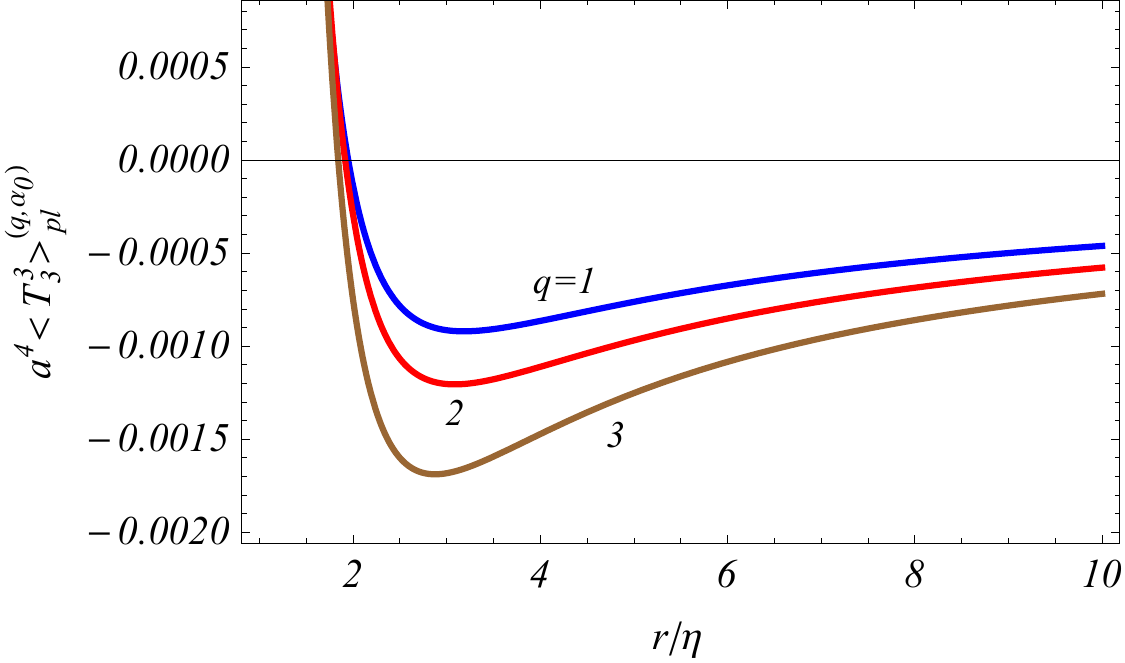}
		\quad
		\includegraphics[scale=0.4]{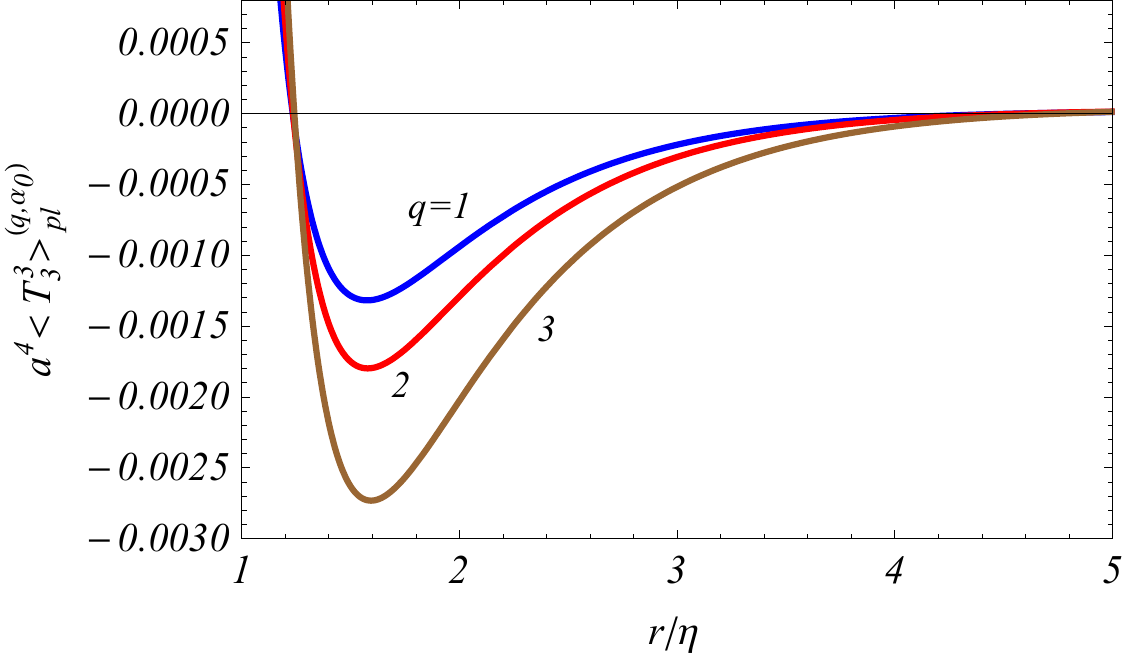}
		\caption{The vacuum normal force on the plate for a minimal coupling and $D=3$ is plotted in terms  $r/\eta$, considering $\alpha_0=0.25$ and $q=1$, 2 and 3. The plot on the left is for $ma=1$, while the plot on the  right is for $ma=2$. The numbers near the curves are the values of $q$.}
		\label{fig5}
	\end{center}
\end{figure}

In Fig. \ref{fig5} the behavior of the vacuum effective pressure on the plate as function of the distance from the string is exhibited for a minimally coupled field, considering $D=3$ and $\alpha_0=0.25$. In the left panel we consider $ma=1$ and in the right one, $ma=2$. The numbers near the curves are values of the angle deficit, $q$.

\section{Conclusion}
\label{conc}
In this paper we have investigated the effects on the bosonic vacuum induced by a flat boundary in a $(1+D)-$dimensional dS background in the presence of cosmic string orthogonal to the plate. Specifically we have calculated the VEVs of the field squared and the energy-momentum tensor. In order to do that we had first to obtain the complete set of normalized positive frequency solutions of the Klein-Gordon equation compatible with the boundary condition imposed on the field at the plate, located at $z=0$. Having obtained this set of solutions, we constructed the Wightman function given by the sum over these modes. This function was expressed as the sum of a contribution associated with the dS space in the presence of a cosmic string without boundary, plus a contribution induced by the plate. Unfortunately  we could not obtain the latter in a closed form for arbitrary values of the $\beta$ parameter present in \eqref{RBC}, so we decided to consider the field obeying  Dirichlet and Neumann boundary conditions separately in our analysis. The corresponding formulas for the VEVs of the field squared and the energy-momentum tensor  differ only by an overall sign.  Because the analysis of the VEVs associated with real scalar fields in higher-dimensional dS in the presence of a cosmic string spacetime have been developed in the literature, in this paper we we focused our investigation on the contribution induced by the boundary. 

In section \ref{phi_2}, we have developed the calculation of the VEV of the field squared induced by the flat boundary, $\langle |\varphi|^2\rangle_{\rm{pl}}$. There we have shown that this quantity is decomposed in two parts, the first,  $\langle |\varphi|^2\rangle_{\rm{pl}}^{(0)}$, corresponds to the contribution induced by the plate in absence of the cosmic string, and the second induced by the cosmic string considering the interaction of the charged bosonic field with the magnetic flux,  $\langle |\varphi|^2\rangle_{\rm{pl}}^{(q,\alpha_0)}$. Because the contribution induced purely by the plate has been investigated in the literature \cite{Saharian:2009ii}, we focus only on the second part. In this way we have investigated this observable in some asymptotic regions of the space. For $r\gg\eta, \ z$, we found \eqref{FSb3}; on the other hand for $z\gg \eta, \ r$, the asymptotic expression is given by \eqref{FSb4}. We can see that both results go to zero with the inverse of the corresponding proper distance at the power $D-2\nu$. In the Minkowskian limit, i.e, considering $a\to\infty$ and fixing the value of $t$, the corresponding VEV was provided in \eqref{FS-M}. Moreover, two figures exhibit the behavior of $\langle |\varphi|^2\rangle_{\rm{pl}}^{(q,\alpha_0)}$, as function of $r/\eta$ and $z/\eta$ for different values of the parameter $q$ and $ma$. They are Figures \eqref{fig1} and \eqref{fig2}, respectively.  

In section \ref{EMT}, we have calculated all components of the VEV of the energy-momentum tensor, $\langle T_{\nu}^{\mu}\rangle_{\rm{pl}}$, induced by the flat plate in dS spacetime in the presence of a cosmic string. We have obtained, besides diagonal components, three off-diagonal ones.\footnote{In the case of absence of a flat boundary anlyzed in \cite{Mello_09}, only one off-diagonal component takes place.} As in the case of VEV of the field squared, this observable is decomposed in a part induced by the boundary only in dS, $\langle T_{\nu}^{\mu}\rangle_{\rm{pl}}^{(0)}$, plus the part induced by the presence of the cosmic string, $\langle T_{\nu}^{\mu}\rangle_{\rm{pl}}^{(q,\alpha_0)}$. Here in this section we analyze only the part induced in the presence of the string. In fact this contribution takes into account, besides the presence of the flat plate in dS, the interaction of the charged scalar field with the magnetic flux running along the string's core  and with the non-trivial topology of the space. As to the magnetic interaction, it provides corrections in the two terms in \eqref{EM-Plate}. The first one associated with the sum over the $k$, and the other associated with the integral. Moreover, we observe that both corrections depend only on the fractional part of the ratio  of the magnetic flux by the quantum one, $\alpha_0$. So the correction corresponds to a Aharonov-Bohm  effect. 
In this section we have proved that the VEV of the energy-momentum tensor satisfies the trace relation, Eq. \eqref{trace}, and the conservation condition, Eq. \eqref{conservation}. Special attention has been given to the energy density, $\langle T_{0}^{0}\rangle_{\rm{pl}}^{(q,\alpha_0)}$. We have investigated its behavior in some limiting cases. For a massless field conformally coupled with the geometry this VEV is given by Eq. \eqref{ED-Brane-massless}. Moreover, the energy density is finite on the string for $z\neq0$, and far from it, $r\gg\eta,z$, its asymptotic behavior is given by \eqref{ED-Asymp-r}. On the surface of the plate, $z=0$, the energy density is finite for $r\neq0$, and in the opposite limit is given by \eqref{ED-Asymp-z}. As we can see in both cases, the energy-density decays with the inverse  of the corresponding proper distance. The Minkowskian limit of the energy-density is given  by \eqref{ED-M}. In order to exhibit the behavior of the energy density as function of $r/\eta$ and $z/\eta$, the proper distances from the string and the plate, respectively, measured in the units of dS curvature radius, $a$, we provide, for $D=3$, the figures \ref{fig3} and \ref{fig4}, considering different values of $q$ and $ma$.  As our final  remark we present the pressure on the flat boundary. This quantity is given by the evaluation of the $(3-3)$ component of the energy-momentum tensor evaluated  at $z=0$. The corresponding expression is given in \eqref{EM33}.  For conformally coupled massless field, the pressure is given by \eqref{presure_mass}. Considering the absence of magnetic flux and for integer $q$, the pressure reads, \eqref{pressure_0}. The numerical behaviors of the pressure, considering $D=3$, and different values of $q$ and $ma$, as function of $r/\eta$ are given in the plots of figure \ref{fig5}.

As we can see, all figures presented in this paper exhibit the behaviours of the corresponding observables considering $\nu$ real ($ma=1$) and $\nu$ imaginary ($ma=2$). From these figures, we can see that the general behaviour of the observables are essentially the same.

As our final remark, we would like to say that the present work may has an important role in the future investigation of vacuum polarization effects in dS spacetime \cite{Elizalde:2010av}. In fact, it is of our particular interest to analyze the Casimir effect induced by a cosmic string orthogonal to two plates in this background geometry.
\section*{Acknowledgments}
W.O.S is supported under grant 2022/2008, Paraíba State Research Foundation (FAPESQ). E.R.B.M is partially supported by CNPq under Grant no 301.783/2019-3.

\end{document}